\documentclass[preprint,review,11pt]{elsarticle}
\usepackage{graphicx,epstopdf}
\usepackage{subfig}
\usepackage{rotating}
\usepackage{amssymb,amsmath}
\usepackage{textcomp}
\usepackage{adjustbox}
\usepackage{tikz}
\usepackage{pgfplots}
\usetikzlibrary{shapes.geometric,arrows.meta,calc}
\pgfplotsset{compat=1.3}%
\usepackage{relsize}
\tikzset{fontscale/.style = {font=\relsize{#1}}
}
\begin{document}

\begin{frontmatter}

\title{An incremental-stencil WENO reconstruction 
for simulation of compressible two-phase flows}

\author[label2]{Bing Wang} 
\author[label2]{Gaoming Xiang}
\author[label2]{Wenbin Zhang} 
\address[label2]{School of Aerospace, Tsinghua University, 100084 Beijing, China}

\author[label1]{Xiangyu Y. Hu\corref{cor1}}
\cortext[cor1]{Corresponding author}
\ead{xiangyu.hu@tum.de}
\address[label1]{Institute of Aerodynamics and Fluid Mechanics, 
Technische Universit\"{a}t M\"{u}nchen, \\ 85748 Garching, Germany }

\begin{abstract}
An incremental-stencil WENO reconstruction method, 
which uses low-order candidate stencils with incrementally increasing width, 
is proposed for finite-volume simulation of 
compressible two-phase flow with the quasi-conservative interface model. 
While recovering the original 5th-order WENO reconstruction in smooth region of the solution, 
due to the presence of 2-point candidate stencils, 
the present method is able to handle closely located discontinuities, 
which is a typical scenario of shock-interface interaction.
Furthermore, a MOOD-type positivity preserving approach is applied to ensure physical meaningful reconstruction. 
Compared with the hybrid method which switches between with the 5th-order WENO 
and 2nd-order reconstructions, 
the present method is free of problem-dependent tunable parameters. 
A number of numerical examples show that the present method 
achieves small numerical dissipation and good robustness for simulating two-phase flow problems 
with strong shock-interface interaction and large density ratio.
\end{abstract}


\end{frontmatter}


\section{Introduction}

\label{sec:intro} The problems of compressible two-phase flow present in
many research fields, such as aero- and astronautics, mechanics, material
science, astrophysics, nuclear engineering even medical sciences. Several
typical examples are underwater explosion, biomedical ultrasound and shock
wave lithotripsy \cite{johnsen2009numerical, lauer2012numerical} and
shock-induced mixing of liquid fuel droplets in scramjet combustor \cite%
{dong2009numerical}, etc. The related researches are valuable from both
theoretical and application points of view.

With the soaring of the computational power and the sliding of computational cost,
numerical simulation becomes one of the main approaches on studying
compressible two-phase flows. There are three main types of simulation
methods according to the underlying Lagrangian, arbitrary
Lagrangian-Eulerian (ALE) and Eulerian meshes on which the two-phase flow
equations are solved. In Lagrangian and ALE methods, the material interface
is tracked naturally by a moving mesh. However, since the mesh deforms with
the flow, the required frequently re-meshing or re-mapping leads to complex
programming and high computational cost. In Eulerian methods, the material
interface can be tracked (front-tracking) by using Lagrangian markers, and
captured (front-capturing) by introducing extra scalars together with
corresponding convection or advection equations. Usually, the extra scalars
can be mass or volume fraction, level set or material-property coefficients,
such as the specific heat ratio. The front-capturing methods can be further
classified into sharp-interface method and smeared interface method. While
in the former the material interface is modeled as sharp surface \cite%
{hirt1981volume, rider1998reconstructing, Fedkiw1999, Hu2006}, in the latter
it is modeled as smooth transition band \citep{niederhaus2008computational,
So2012, ansari2013numerical, coralic2014finite}.

Abgrall et al. \cite{abgrall1996prevent, saurel1999simple,
abgrall2001computations} first proposed an advection equation for the heat
specific ratio. Shyue \cite{shyue1998efficient, shyue1999fluid,
shyue2001fluid} developed several extensions for more complex stiffened-gas,
van der Waals and Mie-Gr\"{u}neison equation of states. One advantage of
these front-capturing methods is that they are able to achieve overall
conservation of mass, momentum and energy. However, since they are not able
to achieve conservation for each phases, these methods are classified as the
quasi-conservative interface model. As a typical smeared-interface model,
the quasi-conservative interface model faces numerical instabilities near
the material interface. It is found that applying 
the characteristic primitive-variable reconstruction 
\cite{coralic2014finite, johnsen2006implementation} other
than the conservative-variable reconstruction, which is usually used in
high-order conservative schemes for simulating single-phase compressible
flows, is able to increase numerical stability considerably. 
With such technique, Johnsen and Colonius \cite{johnsen2009numerical,johnsen2006implementation} is able to apply a 5th-order WENO reconstruction in a
finite-volume formation based on a HLLC Riemann solver for problems of
gas-gas interface interaction and the collapse of air-bubble in water under
moderate shock impact.

However, as will be shown later in this paper, 
even with the characteristic primitive-variable reconstruction, the
5th-order WENO reconstruction still suffers from numerical instability for problems with high density ratio. It is found that, in single-phase flow
simulations, the high-order WENO reconstruction may also suffer from
numerical instability when closely located discontinuities present in the
flow. Such discontinuities may lead to the absence of smooth
candidate stencil for a classical WENO reconstruction \cite%
{titarev2004finite}. Closely located discontinuities are typical scenarios
of shock-interface interaction, in which the shock discontinuity locates
closely with the density discontinuity. The two-phase flow with large density
ratio is prone to such instability since erroneous reconstruction becomes
more serious when the jump of discontinuities increases. Coralic and
Colonius \cite{coralic2014finite} suggested two methods to address such
difficulty. One is the hybridization between the high-order WENO
reconstruction and a 2nd-order reconstruction which is applied only near the
interface. The procedure to solve two-phase problems using a typical hybrid WENO-MUSCL scheme is: firstly, setting the threshold value for scheme selection : $P _{\infty 1}= P _{\infty,  air}$, $P _{\infty 2}= P _{\infty, water}$; secondly, the MUSCL scheme is selected if $P _{\infty 1} < P _\infty < P _{\infty 2}$, otherwise WENO-JS scheme is selected. The limitation of this method is on how to find a general
effective problem-independent interface indicator. Every often such scheme requires different interface detectors for different cases \cite{titarev2004finite, beig2015maintaining}. The other is pre-smoothing the material interface at the initial condition. Besides
increasing the interface thickness artificially, this method may still
suffer the instability problem if the thickness of interface is decreased by highly
stretching flow, such as that happens in high-speed aerobreakup of water
droplet \cite{lasheras1998break}.

In this paper we propose a simple, yet highly efficient incremental-stencil
WENO reconstruction to address the numerical instability of high-order
reconstruction for finite-volume simulation of compressible two-phase flows
with the quasi-conservative interface model. In computing multi-phase flows, the remaining issue of the classical WENO scheme (denoted as WENO-JS) proposed by Jiang and Shu \cite{Jiang1996} is that it has no smooth stencil to choose when there are closely located discontinues due to their large size. Based on the idea of incremental-stencil reconstruction proposed in target ENO scheme \cite{fu2016family}, incremental 2-point and 3-point stencils are used as the candidate stencils for 5th-order WENO reconstruction. Such two-point stencils are used to ensure that the scheme can degenerate to lowest 2nd order when there are closely located discontinues. Furthermore, the present reconstruction is combined with a MOOD-type
positivity preserving method \cite{clain2011high} to ensure physical
meaningful reconstruction. We show by a number of numerical examples on
two-phase flow problems that, while the present reconstruction has good
robustness and is free of tunable interface indicators, it achieves less
numerical dissipation than the hybrid method.

\section{Quasi-conservative interface model}

\label{sec:model} We assume that the fluids are inviscid and compressible,
described by the Euler equations as 
\begin{equation}
\frac{{\partial \mathbf{U}}}{{\partial t}}+\frac{{\partial \mathbf{F}(%
\mathbf{U})}}{{\partial {x_{j}}}}=0.  \label{governing-equation}
\end{equation}%
Here, $t$ and $x_{j}$ are time and dimensions, respectively. $\mathbf{U}%
=(\rho ,\rho {u_{i}},E)^{T}$, and $\mathbf{F}(\mathbf{U})=[\rho {u_{i}},\rho 
{u_{i}}{u_{j}}+p\delta _{ij},(E+p){u_{j}}]^{T}$. This set of equations
describes the conservation laws for mass density $\rho $, momentum density $%
\rho {u_{i}}$ and total energy density $E=\rho e+\rho {u_{i}}{u_{i}}/2$,
where $e$ is the specific internal energy. The relations between density,
internal energy and pressure of both fluids are given by the stiffened-gas
equation of state (EOS) 
\begin{equation}
\rho e=\frac{{p+\gamma {P_{\infty }}}}{{\gamma -1}},  \label{eos}
\end{equation}%
where $\gamma $ is the specific heat ratio, $P_{\infty }$ is a parameter
with the dimension of pressure. Note that the material-properties $\gamma $
and $P_{\infty }$ are different in each fluids. Following the continuous
assumption on pressure and velocity across the material interface, it is
shown that the material properties follow the non-conservative equations 
\cite{abgrall1996prevent, shyue1998efficient, johnsen2006implementation} 
\begin{equation}
\frac{{\partial \boldsymbol{\phi }}}{{\partial t}}+\frac{{\partial \left( 
\boldsymbol{\phi }{u_{j}}\right) }}{{\partial {x_{j}}}}=\boldsymbol{\phi }%
\frac{{\partial {u_{j}}}}{{\partial {x_{j}}}},  \label{material-equation}
\end{equation}%
where 
\begin{equation}
\boldsymbol{\phi }={\left( {{\phi _{1}}\mathrm{,}{\phi _{2}}}\right) ^{T}}={%
\left( {\frac{\gamma }{{\gamma -1}},\frac{{\gamma {P_{\infty }}}}{{\gamma -1}%
}}\right) ^{T}}.  \label{material-properties}
\end{equation}%
With the Euler equations and these non-conservative equations, a
quasi-conservative interface model is defined.

\section{Numerical method}

\label{sec:method} In this paper, a finite-volume method is applied. For
simplicity, Eqs. (\ref{governing-equation}) and (\ref{material-equation})
are assumed in two-dimensions, and the numerical discretization on an
uniform Cartian grid is only presented in the following for the first, i.e. $%
x$ dimension.

\subsection{Semi-discrezization form}

Consider a computational cell ${I_i} = \left[ {{x_i} - \Delta x/2,{x_i} +
\Delta x/2} \right]$, where $\Delta x$ is the grid size, the
semi-discrezization form of the Euler equations is 
\begin{equation}  \label{semi-discrezization-euler}
\frac{{d{{\overline{\mathbf{U}}}_i}}}{{dt}} = - \frac{{{\widehat{\mathbf{F}}%
_{i + 1/2}} - {\widehat{\mathbf{F}}_{i - 1/2}}}}{{\Delta x}}
\end{equation}
where ${\overline{\mathbf{U}}}_i$ represents the cell-averaged conservative
variables, $\widehat{\mathbf{F}}_{i + 1/2}$ and $\widehat{\mathbf{F}}_{i -
1/2}$ represent the numerical fluxes at the right cell face $i + 1/2$ and
the left cell face $i - 1/2$, respectively. The semi-discrezization form of
the material-property equations can be written as 
\begin{equation}  \label{semi-discrezization-material}
\frac{{d{{\overline{\boldsymbol{\phi}}}_i}}}{{dt}} = - \frac{{{\widehat{%
\mathbf{H}}_{i + 1/2}} - {\widehat{\mathbf{H}}_{i - 1/2}}}}{{\Delta x}} + {%
\overline{\boldsymbol{\phi}}_i}\frac{{{u_{i + 1/2}} - {u_{i - 1/2}}}}{{%
\Delta x}}.
\end{equation}
Here, 
\begin{equation}  \label{average-material}
{\overline{\boldsymbol{\phi}}_i} = {\left[ {\overline {\frac{1}{{\gamma - 1}}%
} ,\overline {\frac{{\gamma {P_\infty }}}{{\gamma - 1}}} } \right]_i},
\end{equation}
are the cell-averaged material properties. The first term on the
right-hand-side of Eq. (\ref{semi-discrezization-material}) gives the numerical fluxes for the material properties, and $%
u_{i + 1/2}$ and $u_{i - 1/2}$ in the second term are the flow velocities at
the cell faces.

Once the right-hand side of Eqs. (\ref{semi-discrezization-euler}) and (\ref%
{semi-discrezization-material}) has been evaluated, a time-integration
method, such as the 3rd TVD Runge-Kutta method \cite{Shu1988}, can be
employed to advance the solution in time. Following a general finite volume
method, the numerical fluxes and velocities in Eqs. (\ref%
{semi-discrezization-euler}) and (\ref{semi-discrezization-material}) are
obtained by solving Riemann problems at the cell faces. The initial
conditions for the Riemann problem are reconstructed at the cell face from
the left and right sides by a characteristic primitive-variable
reconstruction.

\subsection{Characteristic primitive-variable reconstruction}

In the characteristic primitive-variable reconstruction \cite%
{johnsen2006implementation}, the $x$-direction components of the Euler and
material-property equations are first rewritten in the primitive-variable
form 
\begin{equation}  \label{primitive-variable-equations}
\frac{{\partial \mathbf{q}}}{{\partial t}} + \mathbf{A} \cdot \frac{{%
\partial \mathbf{q}}}{{\partial x}} = 0,
\end{equation}
where 
\begin{equation}  \label{primitive-variable}
\mathbf{q} = \left( {%
\begin{array}{c}
\rho \\ 
\begin{array}{l}
u \\ 
v%
\end{array}
\\ 
p \\ 
{\frac{1}{{\gamma - 1}}} \\ 
{\frac{{\gamma {P_\infty }}}{{\gamma - 1}}}%
\end{array}%
} \right), \quad \mathbf{A} = \left( {%
\begin{array}{cccccc}
u & \rho & 0 & 0 & 0 & 0 \\ 
0 & u & 0 & {\frac{1}{\rho }} & 0 & 0 \\ 
0 & 0 & u & 0 & 0 & 0 \\ 
0 & {\rho {c^2}} & 0 & u & 0 & 0 \\ 
0 & 0 & 0 & 0 & u & 0 \\ 
0 & 0 & 0 & 0 & 0 & u%
\end{array}%
} \right).
\end{equation}
The characteristic values, and the left (row) and right (column)
eigenvectors of the Jacob matrix for Eq. (\ref{primitive-variable-equations}%
), respectively, are 
\begin{equation}  \label{characteristic-values}
{\lambda _1} = u - a,\quad {\lambda _2} = {\lambda _3} = {\lambda _4} = {%
\lambda _5} = u, \quad {\lambda _6} = u + a.
\end{equation}
and 
\begin{equation}  \label{eigen-vectors}
\mathbf{L} = \left( {%
\begin{array}{cccccc}
0 & {\ - \frac{{\rho c}}{2}} & 0 & {\frac{1}{2}} & 0 & 0 \\ 
1 & 0 & 0 & {\ - \frac{1}{{c^2}}} & 0 & 0 \\ 
0 & 0 & 1 & 0 & 0 & 0 \\ 
0 & 0 & 0 & 0 & 1 & 0 \\ 
0 & 0 & 0 & 0 & 0 & 1 \\ 
0 & {\frac{{\rho c}}{2}} & 0 & {\frac{1}{2}} & 0 & 0%
\end{array}%
} \right), \quad \mathbf{R} = \left( {%
\begin{array}{cccccc}
{\frac{1}{{c^2}}} & 1 & 0 & 0 & 0 & {\frac{1}{{c^2}}} \\ 
{\ - \frac{1}{{\rho c}}} & 0 & 0 & 0 & 0 & {\frac{1}{{\rho c}}} \\ 
0 & 0 & 1 & 0 & 0 & 0 \\ 
1 & 0 & 0 & 0 & 0 & 1 \\ 
0 & 0 & 0 & 1 & 0 & 0 \\ 
0 & 0 & 0 & 0 & 1 & 0%
\end{array}%
} \right).
\end{equation}
Then, a local linearized characteristic decomposition is carried out on the
respective reconstruction stencil to obtain the characteristic variables by
the projection 
\begin{equation}  \label{characteristic-decompsoition}
\widetilde{\mathbf{q}}_j = \mathbf{L}_{i+1/2}\cdot \mathbf{q}_j,
\end{equation}
where $i +r >j >i+1-r$, where $r$ is the radius of the stencil, and $\mathbf{%
L}_{i+1/2}$ is an average between $\mathbf{L}_{i}$ and $\mathbf{L}_{i+1}$.
After that, the left and right values at a cell face for each component of
the characteristic variables, $\widetilde{\mathbf{q}}^{l}_{i+1/2}$ and $%
\widetilde{\mathbf{q}}^{r}_{i+1/2}$, are reconstructed and they are projected
back to obtain the primitive variables at cell face by 
\begin{equation}  \label{back-projected}
\mathbf{q}^{l}_{i+1/2} = \mathbf{R}_{i+1/2}\cdot \widetilde{\mathbf{q}}%
^{l}_{i+1/2}, \quad \mathbf{q}^{r}_{i+1/2} = \mathbf{R}_{i+1/2}\cdot 
\widetilde{\mathbf{q}}^{r}_{i+1/2}.
\end{equation}

\subsubsection{HLLC Riemann solver}

A HLLC type approximate Riemann solver is used since it can sharply resolves discontinuities and is less computational intensive than Roe solver as pointed out by Johnsen and Colonius \cite{johnsen2006implementation}. A brief description of the HLLC approximate Riemann solver is as follows. From the reconstructed primitive variables at cell face $\mathbf{q}%
^{l}_{i+1/2}\equiv\mathbf{q}_{l}$ and $\mathbf{q}^{r}_{i+1/2}\equiv\mathbf{q}%
_{r}$, one can obtain the corresponding conservative and cell-averaged
material variables, represented by $\mathbf{U}_{l}$ and $\mathbf{U}_{r}$,
and the flux functions, represented by $\mathbf{F}_{l}$ and $\mathbf{F}_{r}$%
. The numerical fluxes of HLLC Riemann solver \cite%
{johnsen2006implementation} are given by 
\begin{equation}  \label{hllc-flux}
{\widehat{\mathbf{F}}^{HLLC}} = \frac{{1 + \mathrm{sign}\left( {S^ * }
\right)}}{2}{\mathbf{F}^{ *} _l} + \frac{{1 - \mathrm{sign}\left( {S^ * }
\right)}}{2}{\mathbf{F}^{*} _r}.
\end{equation}
Here, 
\begin{equation}  \label{flux-function}
{\mathbf{F}^{ *}_k} = \frac{{{S^ * }\left( {{S_k}{\mathbf{U}_k} - {\mathbf{F}%
_k}} \right) + {S_k}\left( {{p_k} + {\rho _k}\left( {{S_k} - {u_k}}
\right)\left( {{S^ * } - {u_k}} \right)} \right){\mathbf{D}^*}}}{{{S_k} - {%
S^ * }}}, \quad k = l, r,
\end{equation}
where 
\begin{equation}  \label{wave-speed}
{\mathbf{D}^ * } = \left(0, 1, 0, S^ *, 0, 0\right)^{T}, \quad {S^ * } = 
\frac{{{p_r} - {p_l} + {\rho_l}{u_l}\left( {{S_l} - {u_l}} \right) - {\rho_r}%
{u_r}\left( {{S_r} - {u_r}} \right)}}{{{\rho_l}\left( {{S_l} - {u_l}}
\right) - {\rho_r}\left( {{S_r} - {u_r}} \right)}}
\end{equation}
where $S^ *$ is the middle-wave speed, $S_l$ and $S_r$ represent the left-
and right-wave speeds, respectively, estimated by 
\begin{equation}  \label{wave-speeds}
\begin{array}{ll}
{S_l} = \min \left( {{S^{*}_l},0} \right), & {S_r} = \max \left( {{S^{*}_r},0%
} \right); \\ 
{S^{ * }_l} = \min \left( {\bar u - \bar c,{u_l} - {c_l}} \right), & {S^{ *
}_r} = \max \left( {\bar u + \bar c,{u_r} + {c_r}} \right); \\ 
\overline u = \frac{{\sqrt {{\rho _l}} {u_l} + \sqrt {{\rho _l}} {u_r}}}{{%
\sqrt {{\rho _l}} + \sqrt {{\rho _r}} }}, & \overline {{c^2}} = \frac{{%
\sqrt {{\rho _l}}c_l^2 + \sqrt {{\rho _r}} c_l^2}}{{\sqrt {{\rho _l}} + 
\sqrt {{\rho _r}} }} + \frac{1}{2}\frac{{\sqrt {{\rho _l}} \sqrt {{\rho _r}} 
}}{{{{\left( {\sqrt {{\rho _l}} + \sqrt {{\rho _r}} } \right)}^2}}} {\left( {%
{u_r} - {u_l}} \right)^2}.%
\end{array}%
\end{equation}
Following Ref. \cite{johnsen2006implementation}, the velocity term in Eq. (%
\ref{semi-discrezization-material}) is obtained by 
\begin{equation}  \label{velocities}
{u^{HLLC}} = \frac{{1 + \mathrm{sign}\left( {S^ * } \right)}}{2}{u^{ * }_l}
+ \frac{{1 - \mathrm{sign}\left( {S^ * } \right)}}{2}{u^{ * }_r},\quad {u^{
* }_k} = \frac{{{S^ * }\left( {{S_k} - {u_k}} \right)}}{{{S_k} - {S^ * }}},
\quad k=l, r.
\end{equation}
 
\subsection{Reconstruction method}
\subsubsection{Incremental-stencil WENO (WENO-IS) reconstruction}
\begin{figure}[tbh]
	\begin{center}
		\begin{tikzpicture}[
		dot/.style 2 args={circle,draw=#1,fill=#2,inner sep=2.5pt},
		square/.style 2 args={draw=#1,fill=#2,inner sep=3pt},
		mystar/.style 2 args={star,draw=#1,fill=#2,inner sep=1.5pt},
		mydiamond/.style 2 args={diamond,draw=#1,fill=#2,inner sep=1.5pt},
		scale=0.8
		]
		
		\draw[black,thick]  (0,-1) grid (15,-1);
		\draw[red,thick,yshift=-0.3cm]  (6.25,-2) grid (8.75,-2);
		\draw[red,thick,yshift=-0.3cm]  (3.75,-3) grid (6.25,-3);
		\draw[black,thick,yshift=-0.3cm]  (6.25,-4) grid (11.25,-4);
		\draw[black,thick,yshift=-0.3cm]  (1.25,-5) grid (6.25,-5);
		\foreach \Fila in {1.25,3.75,6.25,8.75,11.25,13.75}{\node[dot={black}{black}] at (\Fila,-1) {};}  
		\foreach \Fila in {5,7.5}{\node[square={black}{white}] at (\Fila,-1) {};}  
		\foreach \Fila in {6.25,8.75}{\node[dot={red}{red}] at (\Fila,-2.3) {};}  
		\foreach \Fila in {3.75,6.25}{\node[dot={red}{red}] at (\Fila,-3.3) {};}  
		\foreach \Fila in {6.25,8.75,11.25}{\node[dot={black}{black}] at (\Fila,-4.3) {};}   
		\foreach \Fila in {1.25,3.75,6.25}{\node[dot={black}{black}] at (\Fila,-5.3) {};}  
		\node[below] at (1.25,-1.3) {$x_{i-2}$};
		\node[below] at (3.75,-1.3) {$x_{i-1}$};
		\node[below] at (6.25,-1.3) {$x_{i}$};
		\node[below] at (8.75,-1.3) {$x_{i+1}$};
		\node[below] at (11.25,-1.3) {$x_{i+2}$};
		\node[below] at (13.75,-1.3) {$x_{i+3}$};
		\node[below] at (5,-1.3) {$x_{i-1/2}$};
		\node[below] at (7.5,-1.3) {$x_{i+1/2}$};
		\node[left]  at (6.25,-2.3) {$S_{1}$};
		\node[left]  at (3.75,-3.3) {$S_{2}$};
		\node[left]  at (6.25,-4.3) {$S_{3}$};
		\node[left]  at (1.25,-5.3) {$S_4$};
		\end{tikzpicture}
	    \end{center}		
		\caption{Full stencil and candidate stencils for the incremental-stencil WENO reconstruction of $\protect\widetilde{q}^{l}_{i+1/2}$.}
		\label{stencil}	
\end{figure}
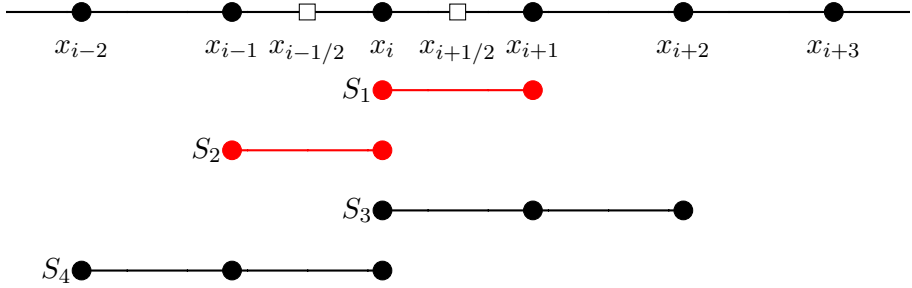

Based on the idea of incremental-stencil reconstruction proposed in the target ENO scheme \cite{fu2016family}, we introduce a new stencil construction approach, the full 5-point stencil is constructed from small stencils with incremental sizes, as shown in Fig. \ref{stencil}. 
For a given number of nodes $r=2$ or $3$, there is a pair of candidate stencils numbered as $S_{2r-1}$ or $S_{2r}$, according to whether its another end node is in the downwind (left) or upwind (right) direction, and all the candidate stencils have one end node at $x_i$. As shown in Fig. \ref{stencil}, the 5th-order WENO-IS reconstruction uses the same upwind-biased full stencil as the classical 5th-order WENO-JS reconstruction \cite{Jiang1996}. The difference is that one of the original 3-point candidate stencil is split into two 2-point stencils. Such design of incremental 2- and 3-point stencils suggests that the present WENO-IS scheme is able to choose one of the 2-point stencil when each candidate stencil of the original WENO-JS reconstruction is crossed by a discontinuity, i.e. there are closely located discontinuities. Note that the present candidate stencils are similar to the incremental stencils of the target ENO scheme \cite{fu2016family}. The difference is that the minimum-size stencil here has 2 points other than 3 points. 

The procedure to obtain the proposed WENO-IS scheme based on the finite volume method is described as follows. In the present reconstruction, a characteristic variable, say $\widetilde{q}%
^{l}_{i+1/2}$, is predicted by the weighted average 
\begin{equation}  \label{final-weighting}
\widetilde{q}^{l}_{i+1/2} = \sum_{k}{w_k}\tilde q_{i + 1/2}^{\left( k \right)},
\end{equation}
where $\tilde q_{i + 1/2}^{\left( k \right)}$ and $w_k$, $k=1,2,3,4$, are
the candidate reconstructed values and their non-linear weights. The
candidate reconstructed values are 
\begin{equation}  \label{candidates}
\begin{array}{l}
\tilde q_{i + 1/2}^{\left( 1 \right)} = \frac{1}{2}{{\tilde q}_i} - \frac{1}{%
	2}{{\tilde q}_{i + 1}}, \\ 
\tilde q_{i + 1/2}^{\left( 2 \right)} = - \frac{1}{2}{{\tilde q}_{i - 1}} + 
\frac{3}{2}{{\tilde q}_i}, \\ 
\tilde q_{i + 1/2}^{\left( 3 \right)} = \frac{1}{3}{{\tilde q}_i} + \frac{5}{%
	6}{{\tilde q}_{i + 1}} - \frac{1}{6}{{\tilde q}_{i + 2}}, \\ 
\tilde q_{i + 1/2}^{\left( 4 \right)} = \frac{1}{3}{{\tilde q}_{i - 2}} - 
\frac{7}{6}{{\tilde q}_{i - 1}} + \frac{{11}}{6}{{\tilde q}_i}.%
\end{array}%
\end{equation}

Inspired by Borges et al. \cite{borges2008improved} and Hu et al. \cite{hu2010adaptive}, 
the weights for the $5$th-order WENO scheme are given by
\begin{equation}
\omega_{k} = \frac{\alpha_{k}}{\sum^{4}_{s=1}\alpha_{s}}, 
\quad \alpha_{k} = d_k\left(1 + \frac{\tau_5}{\beta_{k} + \varepsilon} \right)^{q},
\label{weight-new}
\end{equation}
where $q$ is a positive integer which is set as 1, $\tau_5$ is a global reference smoothness indicator. 
Unlike the classical WENO scheme, 
here the WENO adaption can always find the two-point stencils to increase numerical stability 
for the lowest- or 2nd-order approximation due to the incremental stencil construction.  
The optimal weights $d_k$ are $\{\frac{4}{10}, \frac{2}{10},  \frac{3}{10}, \frac{1}{10}\}$.
With $d_k$, Eq. (\ref{final-weighting}) can be rewritten as
\begin{equation}
\tilde{q}^{l}_{i+1/2} = \sum^{4}_{k=1}d_{k}
\tilde{q}_{k, i+1/2}  + \sum^{4}_{k=1}(\omega_{k} - d_{k}) \tilde{q}_{k, i+1/2}, 
\label{error}
\end{equation}
where the first term on the right-hand-side leads to a $5$th-order approximation.
A sufficient condition for the approximation of Eq. (\ref{final-weighting}) to be of $5$th-order is that the second term
in Eq. (\ref{error}) is at least $\mathcal{O}(\Delta x^{6})$, which requires that the non-linear weights in Eq. (\ref{weight-new}) satisfy the inequality
%
\begin{equation}
\frac{\tau_5}{\beta_{k} + \varepsilon} < \mathcal{O}(\Delta x^{6-r})\label{condition},\, k=1,2,3,4.
\end{equation}
%
In the present WENO-IS reconstruction, using the characteristic variable, say $\widetilde{q}^{l}_{i+1/2}$, the smoothness indicators are given by 
\begin{equation}  \label{smoothness-indicator}
\begin{array}{l}
{\beta _1} = {\left( {{{\tilde q}_{i + 1}} - {{\tilde q}_i}} \right)^2}, \\ 
{\beta _2} = {\left( {{{\tilde q}_i} - {{\tilde q}_{i - 1}}} \right)^2}, \\ 
{\beta _3} = \frac{{13}}{{12}}{\left( {{{\tilde q}_i} - 2{{\tilde q}_{i + 1}}
+ {{\tilde q}_{i + 2}}} \right)^2} + \frac{1}{4}{\left( {3{{\tilde q}_i} - 4{%
{\tilde q}_{i + 1}} + {{\tilde q}_{i + 2}}} \right)^2}, \\ 
{\beta _4} = \frac{{13}}{{12}}{\left( {{{\tilde q}_{i - 2}} - 2{{\tilde q}%
_{i - 1}} + {{\tilde q}_i}} \right)^2} + \frac{1}{4}{\left( {{{\tilde q}_{i
- 2}} - 4{{\tilde q}_{i - 1}} + 3{{\tilde q}_i}} \right)^2}.%
\end{array}%
\end{equation}
here, the global reference smoothness indicator is given by 
\begin{equation}  \label{tau-5}
{\tau _5} = \frac{1}{4}{\left( {{{\tilde q}_{i + 2}} - 2{{\tilde q}_{i + 1}}
+ 2{{\tilde q}_{i - 1}} - {{\tilde q}_{i - 2}}} \right)^2} + \frac{{13}}{{12}%
}{({\tilde q_{i + 2}} - 4{\tilde q_{i + 1}} + 6{\tilde q_i} - 4{\tilde q_{i
- 1}} + {\tilde q_{i - 2}})^2},
\end{equation}
which is the high-order component of the full stencil reconstruction \cite{fu2016family}. The Taylor expansion series of the smooth indicator $\beta _k$ and  global smooth indicator $\tau _5$ at $x _i$ are
\begin{equation} \label{Taylor-expansion}
\begin{array}{l}
\beta _1 = {{\tilde q}_i}^{\prime 2}\Delta {x^2} + {{\tilde q}_i}^{\prime}{{\tilde q}_i}^{\prime \prime}\Delta {x^3} + (\frac{1}{4}{{\tilde q}_i}^{\prime \prime 2}+\frac{1}{3}{{\tilde q}_i}^\prime{{\tilde q}_i}^{\prime \prime \prime })\Delta {x}^4+(\frac{1}{12}{{\tilde q}_i}^\prime{{\tilde q}_i}^{\prime \prime \prime \prime}+\frac{1}{6}{{\tilde q}_i}^{\prime \prime}{f_i}^{\prime \prime \prime})\Delta {x^5}+O(\Delta {x^6})\\
\beta _2 = {{\tilde q}_i}^{\prime 2}\Delta {x^2} - {{\tilde q}_i}^{\prime}{f_i}^{\prime \prime}\Delta {x^3} + (\frac{1}{4}{{\tilde q}_i}^{\prime \prime 2}+\frac{1}{3}{{\tilde q}_i}^\prime{{\tilde q}_i}^{\prime \prime \prime })\Delta {x}^4 - (\frac{1}{12}{{\tilde q}_i}^\prime{{\tilde q}_i}^{\prime \prime \prime \prime}+\frac{1}{6}{{\tilde q}_i}^{\prime \prime}{{\tilde q}_i}^{\prime \prime \prime})\Delta {x^5}+O(\Delta {x^6})\\
\beta _3 = {{\tilde q}_i}^{\prime 2}\Delta {x^2} + (\frac{13}{12}{{\tilde q}_i}^{\prime \prime 2} - \frac{2}{3}{{\tilde q}_i}^\prime{{\tilde q}_i}^{\prime \prime \prime })\Delta {x}^4 + (\frac{13}{6}{{\tilde q}_i}^{\prime \prime}{f_i}^{\prime \prime \prime}-\frac{1}{2}{{\tilde q}_i}^{\prime}{{\tilde q}_i}^{\prime \prime \prime \prime})\Delta {x^5}+O(\Delta {x^6})\\
\beta _4 = {{\tilde q}_i}^{\prime 2}\Delta {x^2} + (\frac{13}{12}{{\tilde q}_i}^{\prime \prime 2} - \frac{2}{3}{{\tilde q}_i}^\prime{{\tilde q}_i}^{\prime \prime \prime })\Delta {x}^4 - (\frac{13}{6}{{\tilde q}_i}^{\prime \prime}{{\tilde q}_i}^{\prime \prime \prime}-\frac{1}{2}{{\tilde q}_i}^{\prime}{{\tilde q}_i}^{\prime \prime \prime \prime})\Delta {x^5}+O(\Delta {x^6})\\
\tau _5 = {{{\tilde q}_i}^{\prime \prime \prime }\Delta {x^6}} + {\frac{13}{12}{{\tilde q}_i}^{\prime \prime \prime \prime }\Delta {x^8} + O(\Delta x^{10})}
\end{array}
\end{equation}
Note that, following the same analysis in \cite{hu2010adaptive, fu2016family}, we get
\begin{equation}
   \frac{\tau _{5}}{\beta _{k} + \varepsilon } = O(\Delta {x^{6-r}}), \quad \text{for } r=2 \text{ or } r=3,
\end{equation}
one can find the present WENO-IS reconstruction achieves 5th-order accuracy in smooth region. Also note that, if the two 3-point stencils are discarded, i.e. by setting $d_3 = d_4 = 0$, the reconstruction degenerates into a 3rd-order reconstruction.

\subsubsection{A modification to the weights}
In Eq. (\ref{smoothness-indicator}), the 2-point stencils have a smooth indicator derived from the integral average of the derivative of the linear polynomial, whereas the 3-point stencils have a smooth indicator derived from the integral average of the derivatives of the 2nd order polynomial. In case of critical points, $\beta _1$ or $\beta _2$ will be small in smooth regions but the relative error of the smooth indicator compared to the exact soluton will be large. Thus, the weights is large compared to the corresponding optimal weights of the two-point stencils near critical points in smooth region and the WENO-IS reconstruction is prone to achieve 2nd-order near the critical points. In order to eliminate this error near critical points in smooth region, a modification of the weights is given by
\begin{equation}  \label{weighting-modification}
{w_k} = \frac{{\alpha _k}}{{\sum\nolimits_{s = 1}^4 {\alpha _s} }}, \quad k=1,2,3,4,
\end{equation}
where
\begin{equation*}
\begin{array}{l}
\alpha_1 = {d_k}\left(1 + \frac{\tau _{5}}{{\beta _{1} + \varepsilon }} \cdot \frac{\tau _{5}}{{\beta _{12} + \varepsilon }} \right),\\
\alpha_2 = {d_k}\left(1 + \frac{\tau _{5}}{{\beta _{2} + \varepsilon }} \cdot \frac{\tau _{5}}{{\beta _{12} + \varepsilon }} \right),\\
\alpha_3 = {d_k}\left(1 + \frac{\tau _{5}}{{\beta _{3} + \varepsilon }}\right),\\
\alpha_4 = {d_k}\left(1 + \frac{\tau _{5}}{{\beta _{4} + \varepsilon }}\right)
\end{array}
\end{equation*}  
$\beta _{12}$ is the smooth indicator of stencil $S _{12}={i-1,i,i+2}$, which in the full three point stencil in the WENO-JS scheme \cite{Jiang1996} and it is given by
\begin{equation}
\beta _{12} = \frac{13}{12}{\left( {\tilde q}_{i-1} - 2{{\tilde q}_{i}}+ {\tilde q}_{i + 1} \right)^2} + \frac{1}{4}{\left( {\tilde q}_{i-1} - {\tilde q}_{i + 1} \right)^2},
\end{equation}
The Taylor expansion series is
\begin{equation}
\beta _{12} = {{\tilde q}_i}^{\prime 2}\Delta {x^2} + (\frac{13}{12}{{\tilde q}_i}^{\prime \prime 2} + \frac{1}{3}{{\tilde q}_i}^\prime{{\tilde q}_i}^{\prime \prime \prime })\Delta {x}^4 +O(\Delta {x^6}),
\end{equation}
Therefore, 
\begin{equation}
\frac{\tau _{5}}{\beta _{12} + \varepsilon }=O(\Delta {x^4}),
\end{equation}
which will not degenerate the order of the reconstruction. 

\subsection{MOOD-type positivity preserving}

It is known that when the material properties or states have very large
jumps across the material interface, the high-order reconstruction can be
erroneous and prone to produce nonphysical states, such as negative pressure
or density, or material properties out of its physically meaningful range.
This numerical phenomena can be generalized as the positivity preserving
problem \cite{einfeldt1991godunov, hu2013positivity}.

In the HLLC type Riemann solver, a physical meaningful wave speed requires that the quatities inside the square root be non-negative. For the stiffened EOS, the wave speed is computed from Eq. (\ref{wave-speed}), the positivity preserving means that $\rho$ should be positive and $p+\gamma P_{\infty}$ should be non-negative. Here, we introduce a MOOD-type approach which is based on "a posteriori" detection \cite{clain2011high}.  A simple detector is used to detect when and how many cells use the MOOD-type positivity preserving. The specific procedures
are as follows. First, the reconstructed primitive variables obtained from
the 5th-order WENO-IS scheme are checked. If the positivity is violated, the
reconstruction is redone without the two 3-points stencils. Then the
primitive variables obtained from the 3rd-order reconstruction is checked
again. If the positivity is still violated, the 1st-order upwind
reconstruction is applied, i.e. 
\begin{equation}  \label{upwind}
\widetilde{q}^{l}_{i+1/2} = \tilde q_{i},
\end{equation}
which is positivity preserving by default.

\section{Convergence test}
\subsection{One-dimensional linear wave equation}
Firstly, the one-dimensional test from Hu et al. \cite{hu2010adaptive} is considered to verify whether the present WENO-IS scheme achieves to the formal order for smooth solutions. We consider the linear advection of an one-dimensional Gauss pulse described as 
\begin{equation} \label{1d-linear-wave-equation}
u = e^{{ - 300(x - {x_c})}^2},
\end{equation}
where $x_c = 0.5$. A periodic boundary condition is applied at $x = 0$ and $x = 1$. The final time is $t = 1$, which corresponds to one period. This problem is computed on different grids with $N = 51, 101, 201, 401, 801$ and $1601$ grid points for convergence study. The time step size is chosen as $\Delta t = 0.5\Delta x^{5/3}$, which is small enough to neglect the temporal truncation error. Fig. \ref{figs:convergence}(a) gives the convergence of the $L_1$ and $L_\infty$ error. It is observed that the present WENO-IS scheme achieves the formal order of accuracy and produces less error than the WENO-JS scheme. 

\subsection{Two-dimensional linear wave equation}
\begin{figure}[tbh]
\includegraphics[width=1.2\textwidth]{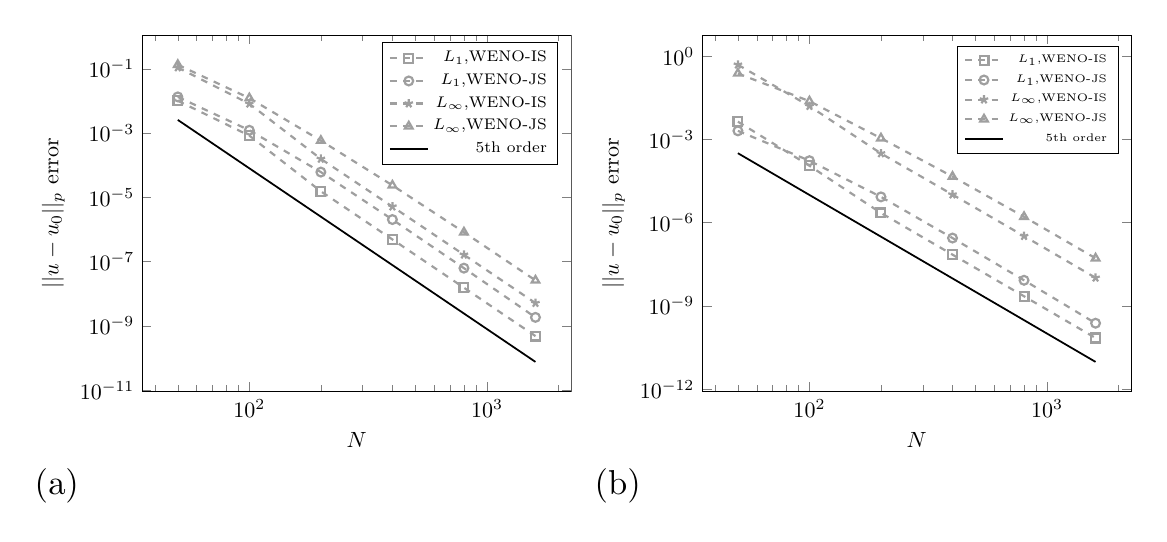}	
\caption{Convergence of the $L_1$ and $L_\infty$ error : (a) one-dimensional linear advection problem; (b) two-dimensional linear advection problem.}
	\label{figs:convergence}
\end{figure}

The two-dimensional test is considered to verify whether the incremental WENO scheme achieves to the formal order for smooth solutions. We consider the linear advection of a two-dimensional Gauss pulse described as 
\begin{equation} \label{2d-linear-wave-equation}
u = e^{ - 300((x - {x_c})^2+(y - {y_c})^2},
\end{equation}
where $(x_c,y_c) = (0.5,0.5)$. The periodic boundary condition is applied for all the boundaries. The final time is $t = 1$, which corresponds to one period. This problem is computed on different grids with $N \times N = 51 \times 51, 101 \times 101, 201 \times 201, 401 \times 401, 801 \times 801$ and $1601 \times 1601$ grid points for convergence study. The time step size is also chosen as $\Delta t = 0.5\Delta x^{5/3}$. Fig. \ref{figs:convergence}b shows the convergence accuracy of the $L_1$ and $L_\infty$ error of the two-dimensional linear advection Gauss pulse problem. It is observed that the present WENO-IS achieves the formal order of accuracy and produces less error better than the WENO-JS scheme.

\section{One-dimensional test problems}

\label{sec:one-dimensional-cases} 

In this section, several one-dimensional
benchmark problems are tested. The following 3 reconstruction methods: the WENO-JS reconstruction, the hybrid WENO-MUSCL reconstruction, which identifies the
material interface region with a user-defined range of $P_\infty$ and the present WENO-IS reconstruction are compared. Note that, for problems with two gases using
ideal-gas EOSs, the hybrid WENO-MUSCL reconstruction and the WENO-JS reconstruction are equivalent because the interface
indicator is not valid due to $P_\infty=0$ for both fluids. As shown by Hu et al. \cite{hu2013positivity} 
and Zhang \& Shu \cite{zhang2012positivity},  independent of numerical scheme, 
a further limited CFL number at least less than 0.5 is required for positivity preserving. For all the test
problems, the number of grid points is 200 and the referenced "exact
solutions" are the MUSCL results computed on a 1600-point grid. The MOOD-type positivity preverving approach is not used for all the one-dimensional test problems. The CFL number for all the one-dimensional test problems is set as 0.5.

\subsection{Gas/liquid interface transportation problem}

This test problem is proposed by Chen and Liang \cite{chen2008flow}. The
liquid phase on the left side and the gas phase on the right side share the
same velocity and pressure, and the gas/liquid interface initially locates
at $x=2$. The initial condition is given as 
\begin{equation}
(\rho, u, p, \gamma, {P_{\infty }}) =\left\{ \begin{array}{lcr}
(1, 100, 1, 7, 3000) & 0\leq x\leq 2, \\ 
(0.001, 100, 1, 1.4, 0) & 2\leq x\leq 4,%
\end{array} \right.  
\label{1d-gas-liquid}
\end{equation}%
The results at $t=0.01$ obtained by all 3 methods, as shown in Fig. \ref%
{figs:1d-gas-liquid-transport}, are in good agreement with the reference
solution. 
\begin{figure}[tbh]
\includegraphics[width=1.2\textwidth]{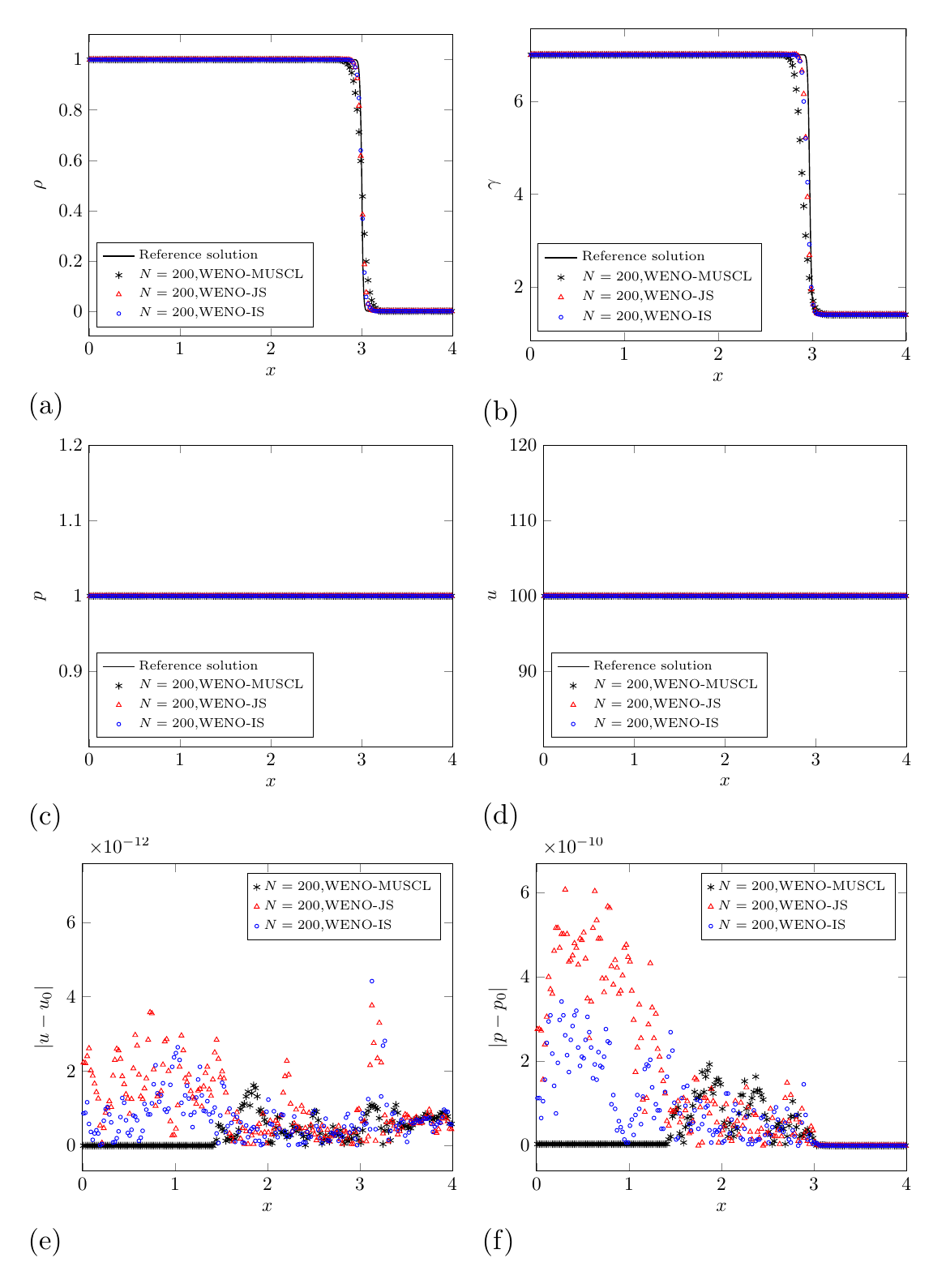}
\caption{Gas/liquid interface transportation problem at $t=0.01$.}
\label{figs:1d-gas-liquid-transport}
\end{figure}
As shown in Figs. \ref{figs:1d-gas-liquid-transport}a and \ref{figs:1d-gas-liquid-transport}b that WENO-IS and WENO-JS produce less numerical dissipation than WENO-MUSCL. Although the hybridization is able to increase numerical stability due to the more stable MUSCL scheme, it introduces excessive numerical dissipation and ruins the accuracy offered by the WENO-JS scheme. It is hard to see the difference from the results obtained by the hybrid scheme and the MUSCL scheme, which is not presented in the paper. The obtained pressure and density profiles in Figs. \ref{figs:1d-gas-liquid-transport}c and \ref{figs:1d-gas-liquid-transport}d suggest that all the numerical solutions are oscillation free.  The errors of $p$ and $u$ shown in Figs. \ref{figs:1d-gas-liquid-transport}e and \ref{figs:1d-gas-liquid-transport}f are really in the scale of $10^{-12}$, i.e. are at the round-off level. Note that, since the profile of $P_{\infty }$ is very
similar as that of $\gamma $, it is not shown here and in the results of other
test problems.

\subsection{Gas/liquid Sod problem}

\begin{figure}[tbh]
\includegraphics[width=1.2\textwidth]{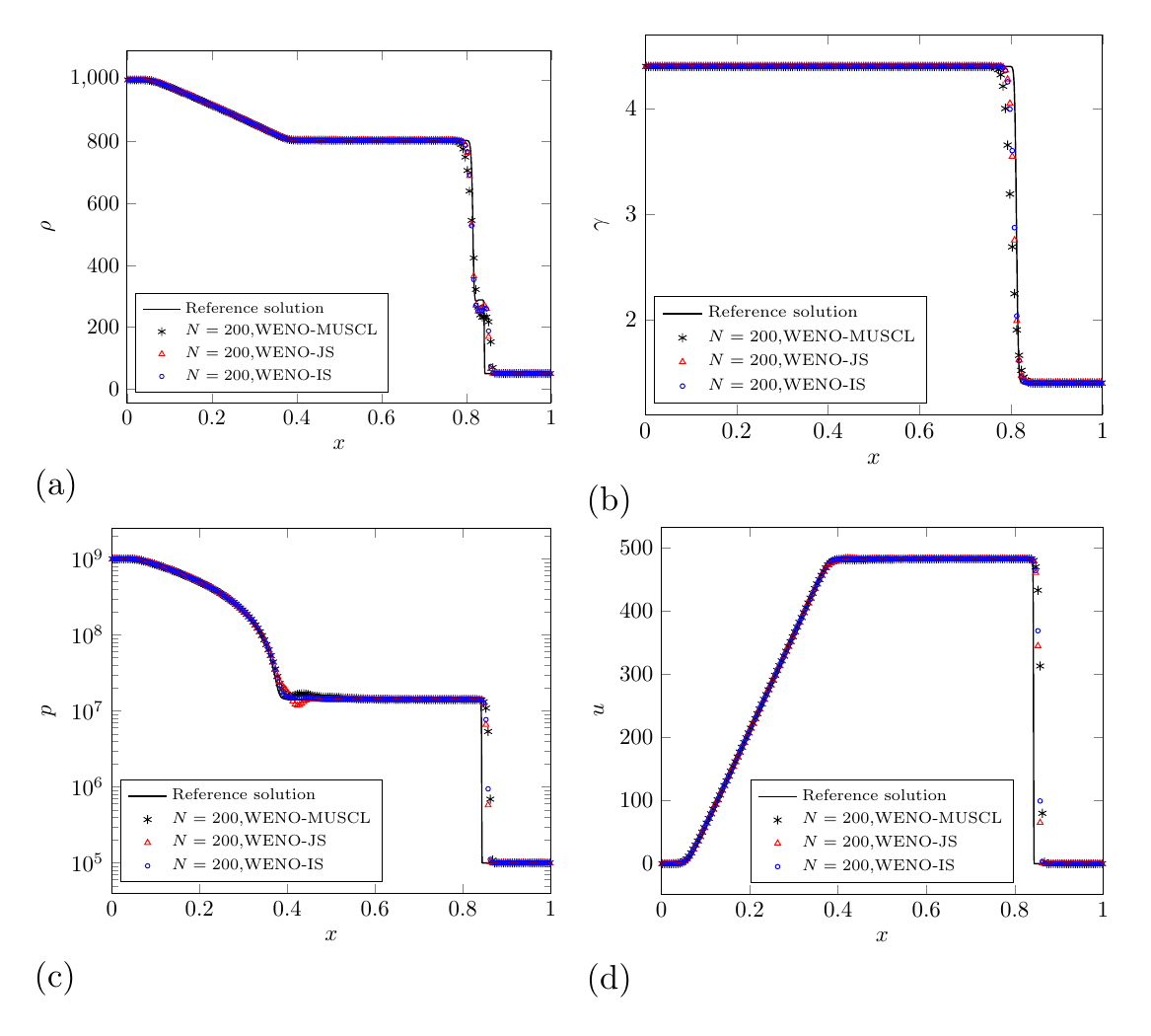}
\caption{Gas/liquid Sod problem at $t=240\textmu $s.}
\label{figs:1d-gas-liquid-sod}
\end{figure}

This gas/liquid Sod problem also comes from Chen and Liang \cite%
{chen2008flow}. The initial discontinuity locates at $x=0.7$ with the high
pressure liquid on the left side and low pressure gas on the right side. The
detailed problem setups are given as 
\begin{equation}
(\rho, u, p, \gamma, P_{\infty }) =\left\lbrace \begin{array}{lcr}
(20, 0, 10^4, 4.4, 6 \times 10^{3}) & 0.0\leq x\leq 0.7, \\ 
(1, 0, 1, 1.4, 0) & 0.7\leq x\leq 1.0,
\end{array} \right.%
\label{1d-gas-liquid-sod}
\end{equation}%
The computational results are given in Fig. \ref{figs:1d-gas-liquid-sod}. 
It can be observed that the numerical results approximate the reference
solution quite well. Note that the pressure profile obtained by the WENO-JS scheme, as
shown in Fig. \ref{figs:1d-gas-liquid-sod}c, exhibits a small overshoot at
the end of rarefaction wave.

\subsection{Shock/bubble interaction problem}

\begin{figure}[tbh]
\includegraphics[width=1.2\textwidth]{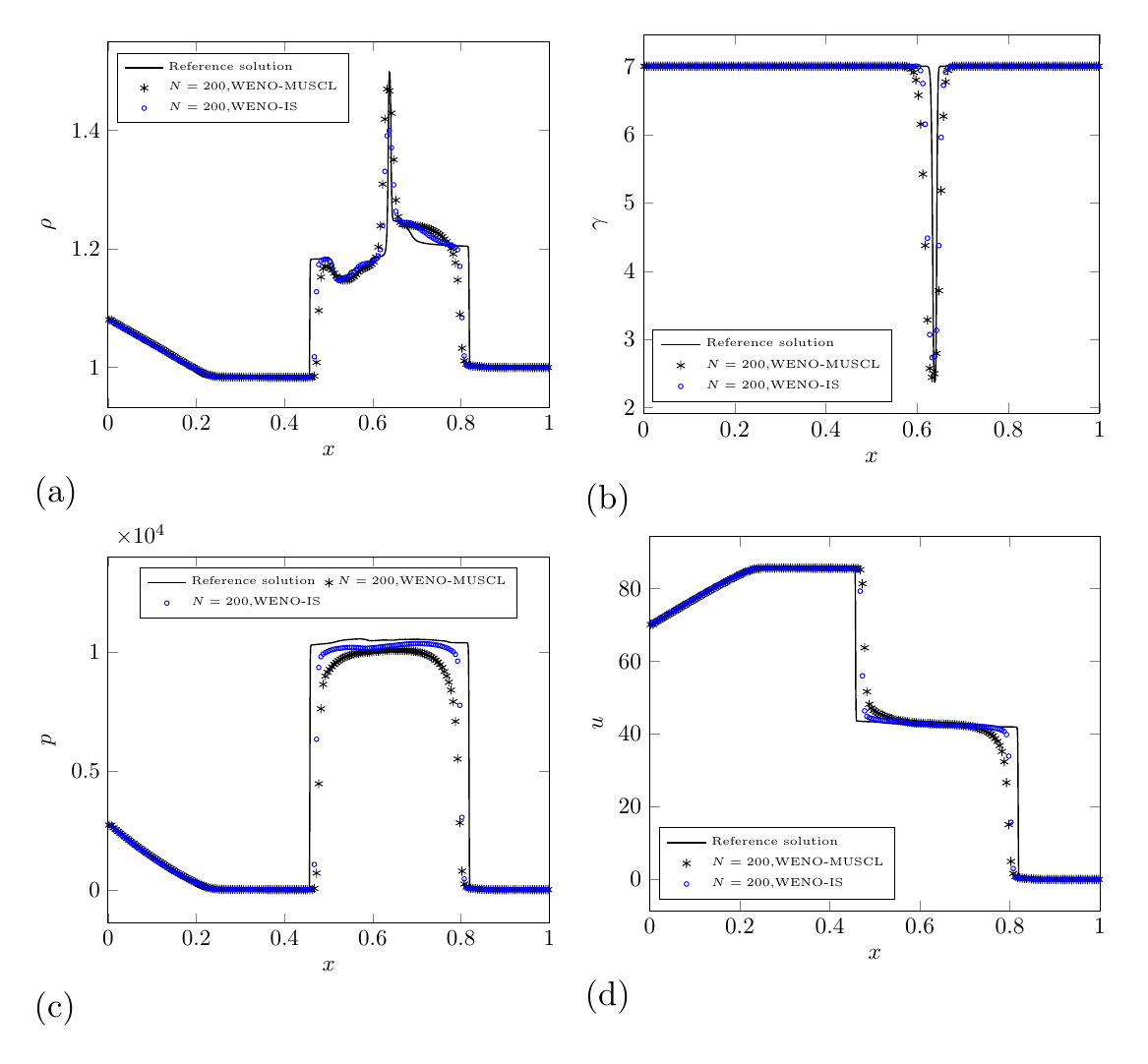}
\caption{Shock/bubble interaction problem at $t=3.87$.}
\label{figs:1d-shock-bubble}
\end{figure}

This test case is a simplified one-dimensional problem on shock/bubble
interaction in liquid. The bubble is in the region $0.4<x<0.6$, and
the shock wave, initially locates at $x=0.25$, impinges the air bubble from
the left side. The problem setup is given as 
\begin{equation}
(\rho, u, p, \gamma, {P_{\infty }}) =\left\{ \begin{array}{llcr}
(1.2199, 42.455, 10^4, 7, 3.31\times 10^3) & 0.0\leq x\leq 0.25, \\ 
(1, 0, 1, 7.0, 3.31\times 10^3) & 0.25\leq x\leq 0.4, \\ 
(10^{-3}, 0, 1.0, 1.4, 0.0) & 0.4\leq x\leq 0.6, \\ 
(1, 0, 1, 7, 3.31\times 10^{3}) & 0.6\leq x\leq 1.0,%
\end{array} \right.  
\label{1d-shock-bubble}
\end{equation}%
Figure \ref{figs:1d-shock-bubble} shows the computational results at time 
$t=3.87$. 
Note that the computation with the WENO-JS reconstruction is not able to run
through this test case. The numerical results show that the shock/bubble
interaction produces complex flow structures, including multiple shock
waves, interfaces and rarefaction waves. It is also observed that the WENO-IS scheme
produces considerably less numerical dissipation than the hyrid WENO-MUSCL scheme.

\subsection{Shock/droplet interaction problem}

\begin{figure}[tbh]
\includegraphics[width=1.2\textwidth]{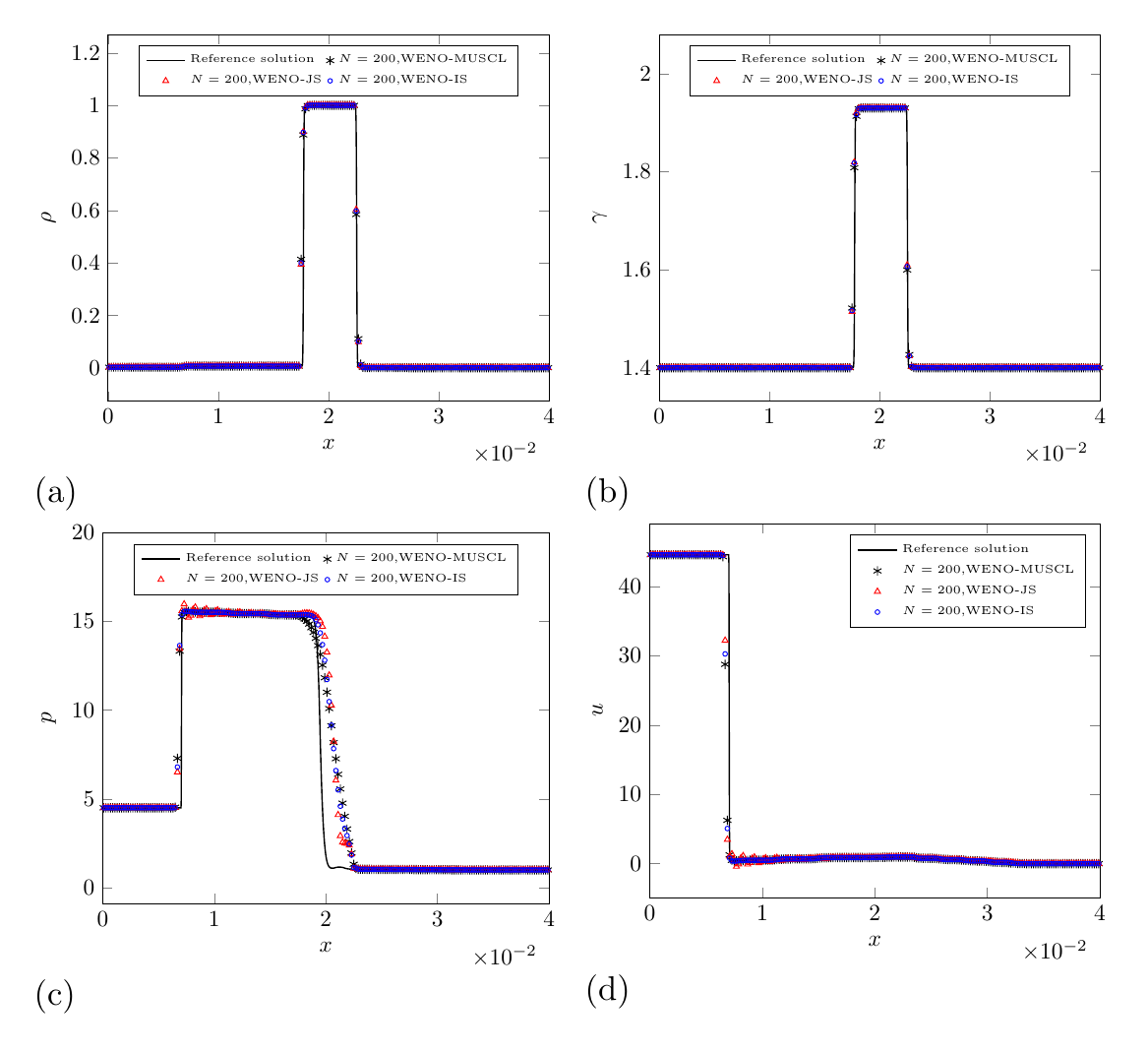}
\caption{Shock/droplet interaction problem at $t=333$.}
\label{figs:1d-shock-droplet}
\end{figure}

This test problem is simplified from the two-dimensional shock/droplet
interaction problem in Chen and Liang \cite{chen2008flow}. An incident
Mach $2$ shock wave initially locates at $x$=$0.016$,
and the droplet is in $0.0176<x<0.0224$ . The shock wave
propagates from left to right. The detailed setup is 
\begin{equation}
(\rho, u, p, \gamma, {P_{\infty }}) =\left\{ \begin{array}{llccrrrr}
(3.2\times 10^{-3},44.59, 4.5,1.4,0) & 0.0\leq x\leq 0.016, \\ 
(1.2\times 10^{-3}, 0, 1, 1.4, 0)  & 0.016\leq x\leq 0.0176, \\ 
(1, 0, 1, 1.932, 1.1645\times 10^4) & 0.0176\leq x\leq 0.0224, \\ 
(1.2\times 10^{-3}, 0, 1, 1.4, 0) & 0.0224\leq x\leq 0.04,%
\end{array} \right.  
\label{1d-shock-droplet}
\end{equation}%
Figure \ref{figs:1d-shock-droplet} shows the results at
time $t=333$. 
As shown in Fig. \ref{figs:1d-shock-droplet}(c) and Fig. \ref{figs:1d-shock-droplet}(d), the WENO-JS scheme produces both pressure and velocity oscillations at the reflected shock front. However, the WENO-IS scheme doesn't produce such spurious oscillation  as shown in Fig. \ref{figs:1d-shock-droplet}(c) and Fig. \ref{figs:1d-shock-droplet}(d). This is due to that the WENO-IS scheme is able to select the two-point stencil and reduces the reconstruction order to suppress non-physical oscillation.

\section{Two-dimensional test problems}

\label{sec:two-dimensional-cases} In this section, simulations of two-dimensional problems involving shock-interface interaction are performed. The first two problems involve single-phase but multiple gaseous components and the other two problems involve air and water phases. For the single phase multi-component test problems, the MOOD-type positivity preserving approach is not activated. The CFL number is set as 0.5 for all the two-dimensional test problems.

\subsection{Single-phase multi-component (air-R22 and air-He) problems}

\label{sec:gas-gas} We consider a shock wave interacting respectively with the helium
(He) or hydrochlorofluorocarbon (R22) gas cylinders, which were first studied
experimentally in Haas and Sturtevant \cite{haas1987interaction} and subsequent numerical studies were implemented in \cite{johnsen2006implementation,quirk1996dynamics}. While the
R22 bubble has higher density than the air, the He bubble has lower density. 
\begin{table}[tbp]
\begin{center}
\begin{tabular}{|c|c|c|c|c|c|c|}
\hline
& \multicolumn{3}{|c|}{Air-R22} & \multicolumn{3}{|c|}{Air-He} \\ \hline
$Ms$ & \multicolumn{6}{|c|}{1.22} \\ \hline
Materials & Air1 & Air2 & R22 & Air1 & Air2 & He \\ \hline
$\rho ($kg$/$m$^{3})$ & 1.4 & 1.927 & 4.415 & 1.4 & 1.927 & 0.255 \\ \hline
$u($m$/$s$)$ & 0 & -114.42 & 0 & 0 & -114.42 & 0 \\ \hline
$p($MPa$)$ & 0.1 & 0.157 & 0.1 & 0.1 & 0.157 & 0.1 \\ \hline
$\gamma $ & \multicolumn{2}{|c|}{1.4} & 1.249 & \multicolumn{2}{|c|}{1.4} & 
1.648 \\ \hline
$P_{\infty }($GPa$)$ & \multicolumn{6}{|c|}{0} \\ \hline
\end{tabular}%
\caption{The setup parameters for single-phase multi-component (air-R22 and
air-He) cases.}
\label{table:initial-condition-air-he-r22}
\end{center}
\end{table}
The setups of the pre- and post-shock wave and the cylinder properties are
shown in Tab. \ref{table:initial-condition-air-he-r22}. 
\begin{figure}
	\begin{center}	
		\begin{tikzpicture}[scale=0.9]
		\draw[black,very thick] (0,0) rectangle (13.35,-4.45);
		\draw[black,very thick] (10.15,0) -- (10.15,-4.45);
		\draw[fill=blue!30!white,very thick] (8.9,-2.225) circle (1.25cm);
		\draw[black,very thick,{Stealth}-{Stealth}]  (0,-5.15) -- (13.35,-5.15);
		\draw[black,very thick,{Stealth}-{Stealth}]  (14,0) -- (14,-4.45);
		\draw[black,very thick,{Stealth}-{Stealth}]  (7.55,-0.975) -- (7.55,-3.475);
		\draw[black,very thick,{Stealth}-{Stealth}]  (10.15,0.25) -- (13.35,0.25);
		\draw[black,very thick,{Stealth}-]  (0,-3.45) -- (0,-4.45);
		\draw[black,very thick,-{Stealth}]  (0,-4.45) -- (1,-4.45);
		\node[above] at (6.675,-5.15) {267};
		\node[above] at (11.75,0.25) {64};
		\node[left] at (14,-2.225) {\rotatebox[]{90}{89}};
		\node[left] at (7.55,-2.225) {\rotatebox[]{90}{50}};
		\node[below] at (1,-4.45) {x};
		\node[left] at (0,-3.45) {y};
		\node[right] at (1,-2.225) {Air1};
		\node[] at (8.9,-2.225) {R22/He};
		\node[right] at (11,-2.225) {Air2};
		\end{tikzpicture}
		\caption{Schematic diagram of the computational domain (mm) for	the single-phase multi-component (air-R22 and air-He) cases.}
		\label{figs:2d-single-phase-setup}	
	\end{center}
\end{figure}
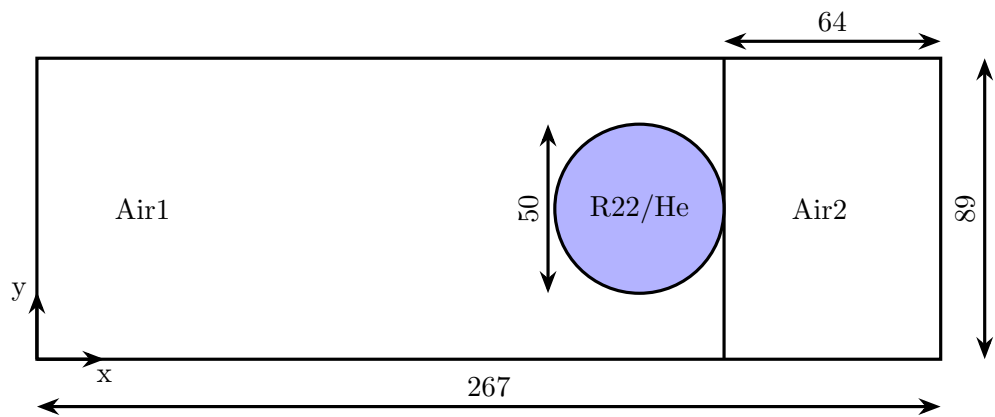
The initial setups as well as the geometries of the computational domain are shown in Fig. \ref%
{figs:2d-single-phase-setup}. Reflected boundary condition is applied at the
upper and lower walls, and constant extrapolation at the left and right
boundaries. A $3000\times 1000$ grid is used for both air-R22 and air-He
problems.

The Schlieren images for the air-R22 problem are shown in Fig. \ref%
{figs:2d-single-phase-air-r22}. The left column gives the experimental
results from Haas and Sturtevant \cite{haas1987interaction} and the right
column gives the present numerical results. 
\begin{figure}[t]
\includegraphics[width=0.75\textwidth]{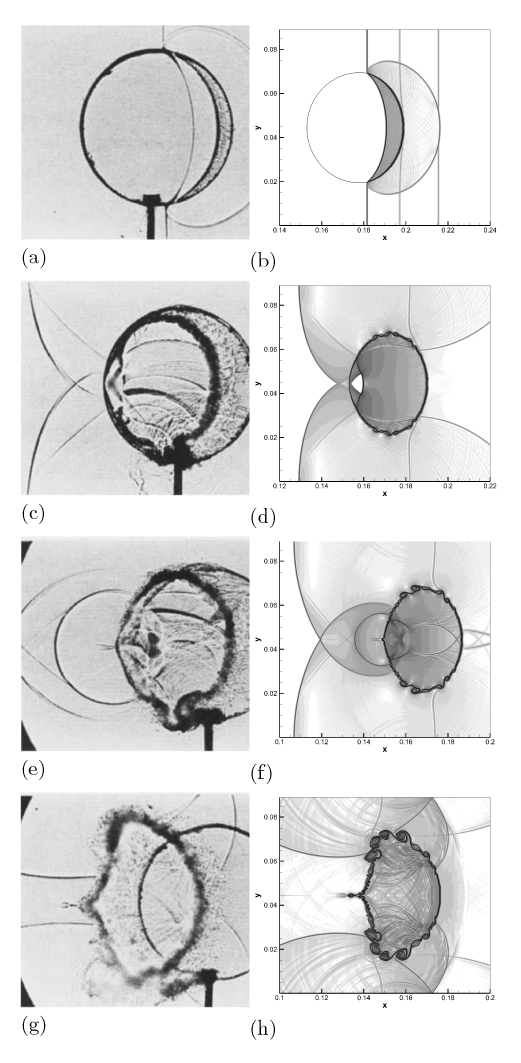}
\caption{Air-R22 case: Experimental (left) and numerical (right) Schlieren images 
	at different time instances: $t=55$, 190, 250, 420 \textmu s. 
	Note that there are small density disturbances 
	behind the incident shock due to the initial setup. 
	Such small numerical disturbances can be seen from 
	all the two-dimensional simulations in this paper, 
	and many previous simulations, such as those in Refs. 
	\cite{johnsen2006implementation, chang2007robust, Hu2009, hawker2012interaction}. }
\label{figs:2d-single-phase-air-r22}
\end{figure}
It is clearly shown that the numerical results agree well with the
experimental results, such as the shock waves and the deformation of the R22
bubble. We can see the complex interactions between the transmitted,
reflected, diffracted and refracted shocks after the incident shock impinges the bubble wall. Note that the Kelvin-Helmholtz instability
develops along the air-R22 interface and the small rolling up structures
obtained here agree well with the result from So et al. \cite%
{So2012} (their Fig. 6) using an interface sharpening technique in their simulation.

The Schlieren images for the air-He problem are shown in Fig. \ref%
{figs:2d-single-phase-air-helium}. 
\begin{figure}[tbh]
\includegraphics[width=0.75\textwidth]{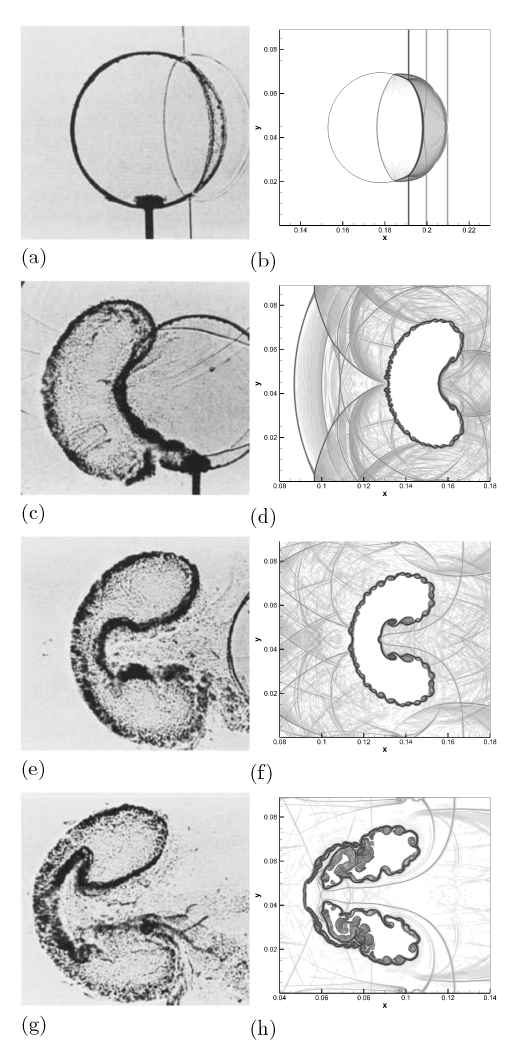}
\caption{Air-He case: Experimental (left) and numerical (right) Schlieren images
	at different time instances: $t=30$, 245, 380, 674 \textmu s.}
\label{figs:2d-single-phase-air-helium}
\end{figure}
Again, the left column gives the experimental results from Haas and
Sturtevant \cite{haas1987interaction} and the right column is the present
numerical results. The results from our nemerical scheme are in good
agreement with that of the experimental results in qualitative aspect. The
incident shock is transmitted and reflected when it contacts with the He
bubble. The upstream bubble wall is compressed and the tranverse
jet forms. Note that the air-He interface rolls up, similar to the air-R22
case, due to the Kelvin-Helmholtz instability. These rolling-up
structures are consistent with the numerical results (their Fig. 12) from
Johnsen and Colonius \cite{johnsen2006implementation} computed with WENO
reconstruction on a coarser grid.

\begin{figure}[tbh]
\includegraphics[width=\textwidth]{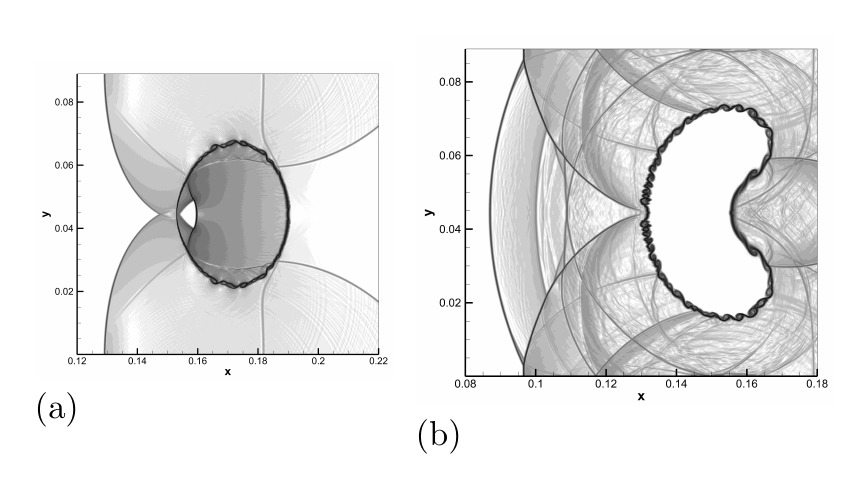}
	\caption{Numerical schlieren using the classical WENO-JS scheme: (a) Air-R22 case, $t=190$ \textmu s; (b) Air-He case, $t=245$ \textmu s.}
	\label{figs:2d-single-phase-weno-js}
\end{figure}

As has been stated in Section \ref{sec:one-dimensional-cases}, for the air-R22 problem and the air-He problem, WENO-MUSCL and WENO-JS are equivalent because the interface indicator is not valid due to $P_\infty=0$. Thus, the numerical results of the present WENO-IS scheme are only compared with the classical WENO-JS scheme, as shown in Fig. \ref{figs:2d-single-phase-weno-js}. For the air-R22 problem, Fig. \ref{figs:2d-single-phase-weno-js}a shows the time instance that the incident shock passes through the downstream R22 bubble wall. Near the downstream R22 bubble wall, a triangle region is formed due to the transmittion of the incident shock from the upstream wall and downstream wall of the R22 bubble. Two slip lines close to the the upper and lower wall of the R22 bubble are also seen, which are due to the reflection of the transmitted shock wave at the R22 bubble wall. These wave structures are well captured by using both the WENO-JS scheme and the WENO-IS scheme. For the air-He problem, Fig. \ref{figs:2d-single-phase-weno-js}b shows the time instance that the incident shock intersects at the downstream wall of the He bubble and there are complex wave structures due to reflection at the upper and lower boundary. These wave structures are similar to the results using the WENO-IS scheme shown in Fig. \ref{figs:2d-single-phase-air-helium}(d). But, for both the air-R22 problem and the air-He problem, the interface between air and R22 or He bubble is a bit more smeared using the WENO-JS scheme when compared with the present WENO-IS scheme.

\subsection{Two-phase test (air-water) problems}

In this section, two problems, i.e. the shock wave interacting with a droplet
in air and an air-bubble in liquid are considered. Compared with the single-phase
multi-component problems in Sec. \ref{sec:gas-gas}, the numerical simulations
of these two problems require very stable and robust numerical method due to
the high density ratio and very strong shock-interface interaction.

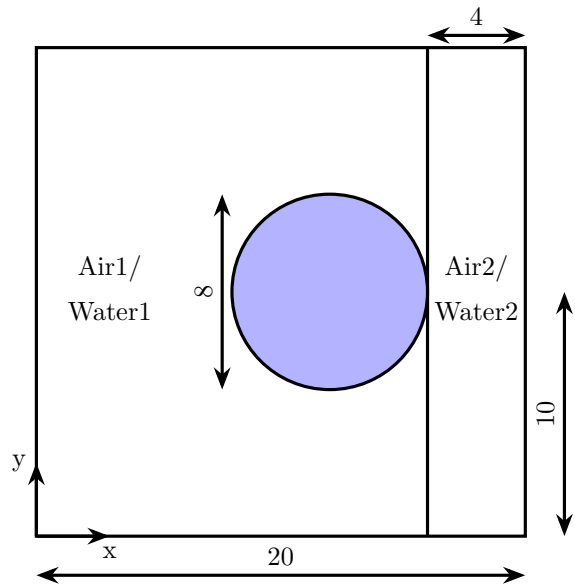
\begin{figure}[tbh]
	\begin{center}
		\begin{tikzpicture}[scale=0.65,font=\small]
		\draw[black,very thick] (0,0) rectangle (10,-10) ;
		\draw[black,very thick] (8,0) -- (8,-10);
		\draw[fill=blue!30!white,very thick] (6,-5) circle (2cm);
		\draw[black,very thick,{Stealth}-{Stealth}]  (0,-10.8) -- (10,-10.8);
		\draw[black,very thick,{Stealth}-{Stealth}]  (10.8,-5) -- (10.8,-10);
		\draw[black,very thick,{Stealth}-{Stealth}]  (3.8,-3) -- (3.8,-7);
		\draw[black,very thick,{Stealth}-{Stealth}]  (8,0.25) -- (10,0.25);
		\draw[black,very thick,{Stealth}-]  (0,-8.5) -- (0,-10);
		\draw[black,very thick,-{Stealth}]  (0,-10) -- (1.5,-10);
		\node[above] at (5,-10.8) {20};
		\node[above] at (9,0.25) {4};
		\node[left] at (10.8,-7.5) {\rotatebox[]{90}{10}};
		\node[left] at (3.8,-5) {\rotatebox[]{90}{8}};
		\node[below] at (1.5,-10) {x};
		\node[left] at (0,-8.5) {y};
		\node[above] at (1.5,-5) {Air1/};
		\node[below] at (1.5,-5) {Water1};
		\node[above] at (9,-5) {Air2/};
		\node[below] at (9,-5) {Water2};
		\end{tikzpicture}
		\caption{Schematic diagram of the computational domain (mm) for the two-phase (air-water) test cases.}
		\label{figs:2d-air-water-setup}
	\end{center}
\end{figure}

The initial setup and boundary conditions are shown in Tab. \ref%
{table:initial-condition-air-water}. 
\begin{table}[tbp]
\begin{center}
\begin{tabular}{|c|c|c|c|c|c|c|}
\hline
& \multicolumn{3}{|c|}{Air-Water} & \multicolumn{3}{|c|}{Water-Air} \\ \hline
$Ms$ & \multicolumn{3}{|c|}{2.0} & \multicolumn{3}{|c|}{1.547} \\ \hline
Materials & Air1 & Air2 & Water & Water1 & Water2 & Air \\ \hline
$\rho ($kg$/$m$^{3})$ & 1.2 & 3.2 & 1000 & 1000 & 1219.9 & 1.0 \\ \hline
$u($m$/$s$)$ & 0 & -434 & 0 & 0 & -424.55 & 0 \\ \hline
$p($MPa$)$ & 0.1 & 0.456 & 0.1 & 0.1 & 1000 & 0.1 \\ \hline
$\gamma $ & \multicolumn{2}{|c|}{1.4} & 4.34 & \multicolumn{2}{|c|}{7.0} & 
1.4 \\ \hline
$P_{\infty }($GPa$)$ & \multicolumn{2}{|c|}{0} & 0.484 & 
\multicolumn{2}{|c|}{0.331} & 0 \\ \hline
\end{tabular}%
\caption{The setup parameters for air-water cases.}
\label{table:initial-condition-air-water}
\end{center}
\end{table}
The computational domain is shown in Fig. \ref{figs:2d-air-water-setup}. 
The non-reflection boundary conditions are applied at all the domain
boundaries. In order to study the convergence property, several grid resolutions, up to $800\times 800$ for the shock droplet case and $1600\times 1600$ for shock-bubble case, are used for the simulations. For the former case, the MOOD-type positivity preserving is not activated for the present WENO-IS scheme.

\begin{figure}[tbh]
\includegraphics[width=\textwidth]{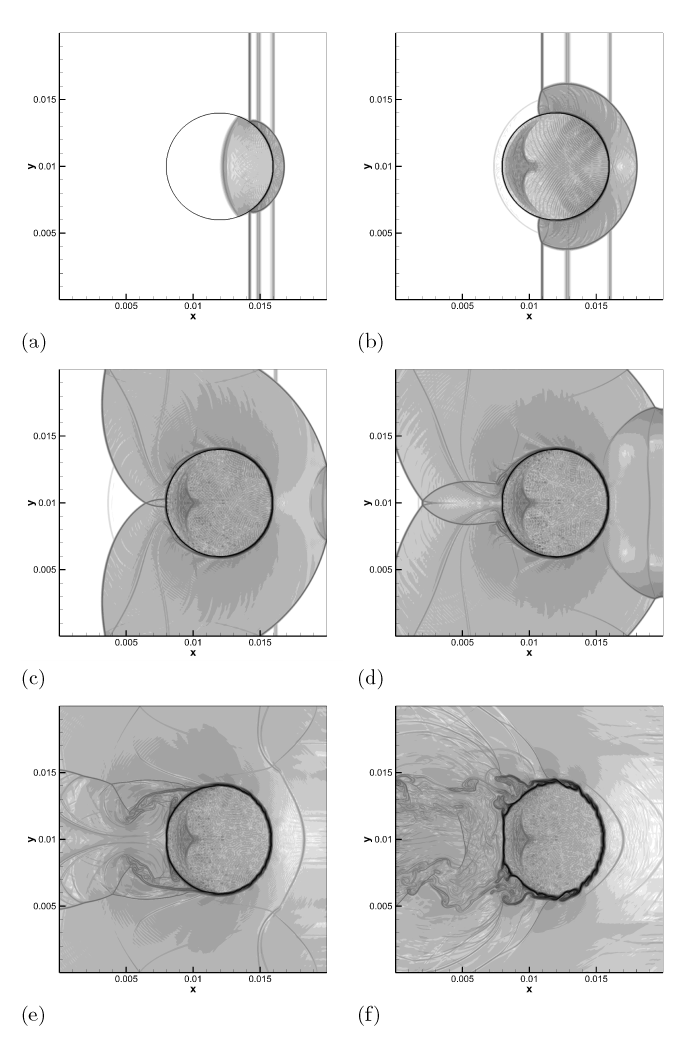}
\caption{Numerical Schlieren images of the shock-droplet case at
	different time instances: $t=2.41$, $7.15$, $17.97$, $23.85$, $36.4$, $71.67$%
	\textmu s.}
\label{figs:2d-air-water-droplet}
\end{figure}

The numerical Schlieren images for the water-droplet problem at different
time instances are shown in Fig. \ref{figs:2d-air-water-droplet}. It is observed
that the early stage of the overall process is in good agreement with
previous observations \cite{wierzba1988experimental, igra2001numerical}. The
differences in the later stage are not unexpected since the present
simulation is two-dimensional and neglects viscous effects and surface
tension. 
After the incident shock impinges at the droplet, the transmitted wave
reflects and impinges at the droplet, the transmitted wave arrives at the downstream surface much
earlier than the diffraction waves because the sound speed is much faster in
the water than that in the air. Later, the secondary
transmitted shock wave forms, though it is very weak compared to the
incident wave. A Mach-reflection structure is produced near the upper and
lower surface, and the rarefaction waves within the droplet produce complex
patterns. Some microseconds later, parts of the liquid are stripped out from
the droplet. This is mainly due to the stripping effect, which is a main mechanism in aerobreakup \cite{lasheras1998break}.

\begin{figure}[tbh]
\includegraphics[width=\textwidth]{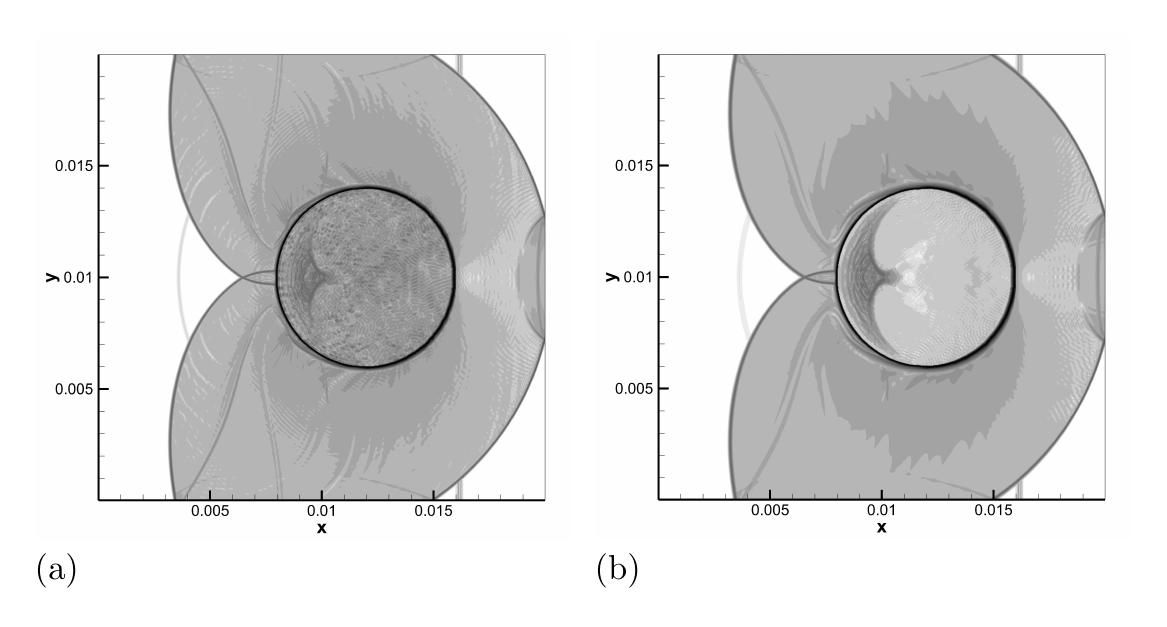}
	\caption{Numerical Schlieren images of the shock-droplet case at time instance $t=17.97$ \textmu s: (a) the classical WENO-JS scheme; (b) the hybrid WENO-MUSCL scheme.}
	\label{figs:2d-air-water-droplet-weno-js-muscl}
\end{figure}

The numerical schlierens obtained by using the WENO-JS scheme and the hybrid WENO-MUSCL scheme are also shown in Fig. \ref{figs:2d-air-water-droplet-weno-js-muscl}. Note that the WENO-JS scheme requires the MOOD-type positivity preserving approach be activated for successful computation. The time instance is $t=17.97$ \textmu s, which corresponds to that of Fig. \ref{figs:2d-air-water-droplet}c. At this time, the incident shock intersect near the downstream droplet surface and gradually overlap the re-transmitted shock in the remaining air region. It is observed that not only the two-phase interface, but also the wave structures include the complicated reflected rarefaction wave inside the droplet and the re-transmitted shock wave as well as the slip lines using WENO-MUSCL scheme are smeared, which indicates that WENO-MUSCL has larger dissipation than the present WENO-IS scheme. 

\begin{figure}[tbh]
\includegraphics[width=\textwidth]{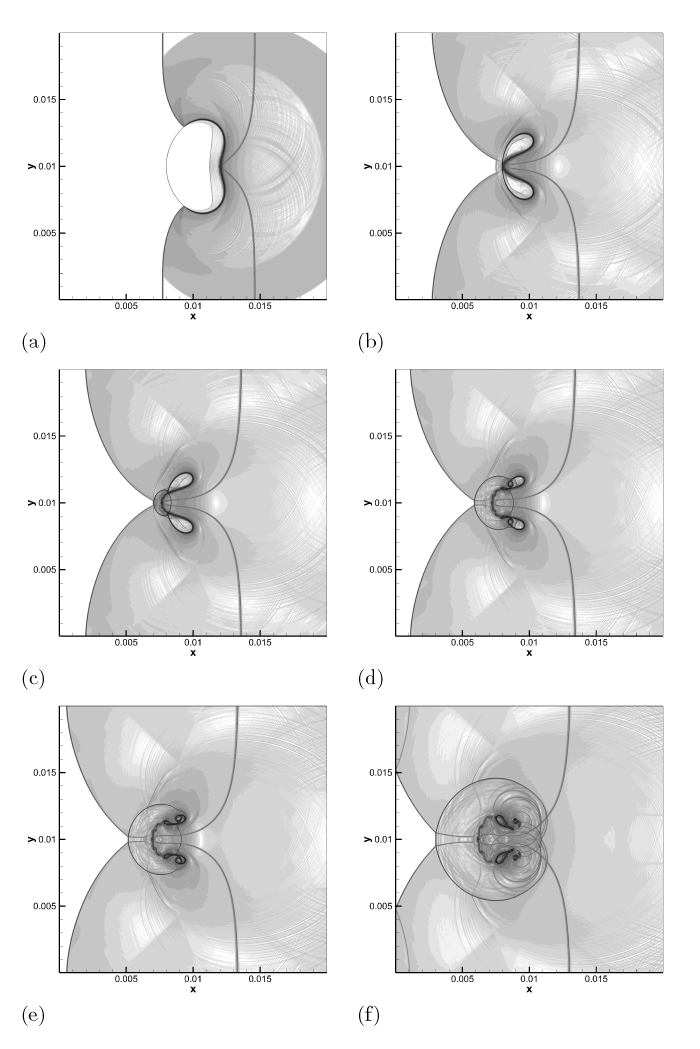}
\caption{Numerical Schlieren images of the shock-bubble case at different time instances: $t=3.3$, $5.4$, $5.7$, $6.1$, $6.4$, $7.1$ \textmu s.}
\label{figs:2d-air-water-bubble}
\end{figure}

For the problem of two-phase shock-bubble interaction in liquid, the MOOD-type positivity preserving approach is activated for both the WENO-IS scheme and the WENO-JS scheme. 
The numerical Schlieren images for the shock-bubble case at different
instances are shown in Fig. \ref{figs:2d-air-water-bubble}. It is observed
that the overall evolution of the shock-bubble interaction is consistent
with the experimental and numerical results in Refs. \cite%
{chang2007robust, hawker2012interaction,wierzba1988experimental}. 
At the early stage, it is obvious that the bubble gets compressed and
deforms when it is impacted by the shock. As shown in Figs. \ref%
{figs:2d-air-water-bubble}(a) and \ref{figs:2d-air-water-bubble}(b), while
the diffract shock propagates faster along the bubble surface, the primary
jet forms, impacts the downstream surface and splits the bubble in two
parts. Such impact produces a very strong water-hammer shock wave, as shown
in Figs. \ref{figs:2d-air-water-bubble}(c) and \ref{figs:2d-air-water-bubble}%
(d), and each split bubble part continues to be compressed and the secondary
jet is formed. Finally, the bubble is divided into four parts and the shock
wave structures become complex and interacting with each other, as shown in
Figs. \ref{figs:2d-air-water-bubble}(e) and \ref{figs:2d-air-water-bubble}(f).

\begin{figure}[tbh]
\includegraphics[width=\textwidth]{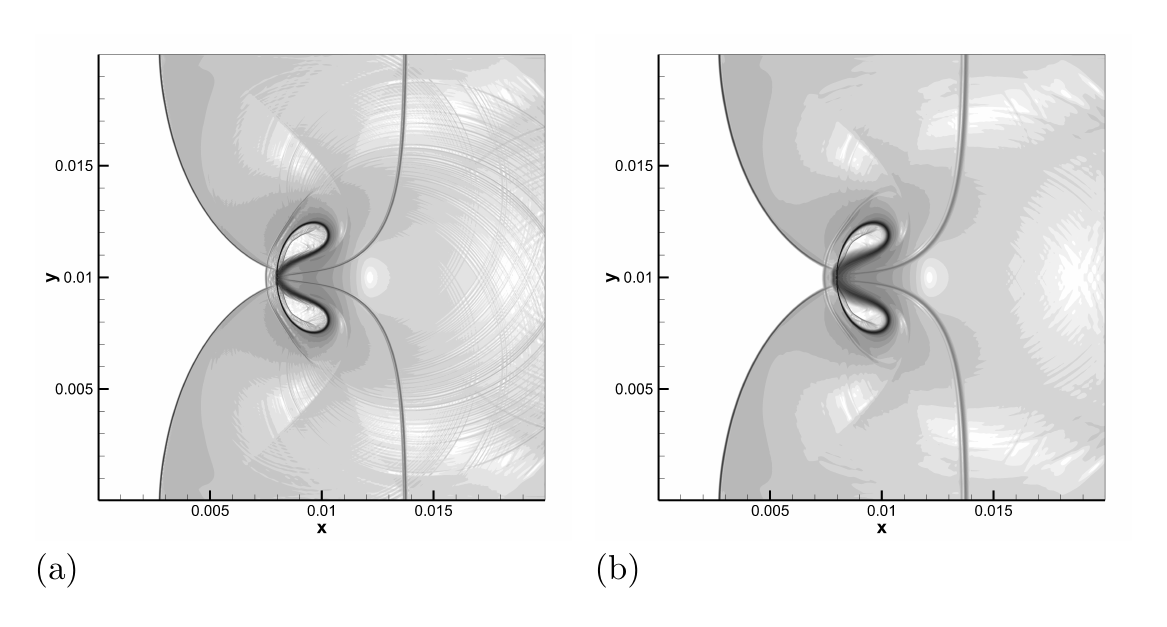}
	\caption{Numerical Schlieren images of the shock-bubble case at time instance $t=5.4$ \textmu s: (a) the classical WENO-JS scheme; (b) the hybrid WENO-MUSCL scheme.}
	\label{figs:2d-air-water-bubble-weno-js-muscl}
\end{figure}

\begin{figure}[tbh]
\includegraphics[width=\textwidth]{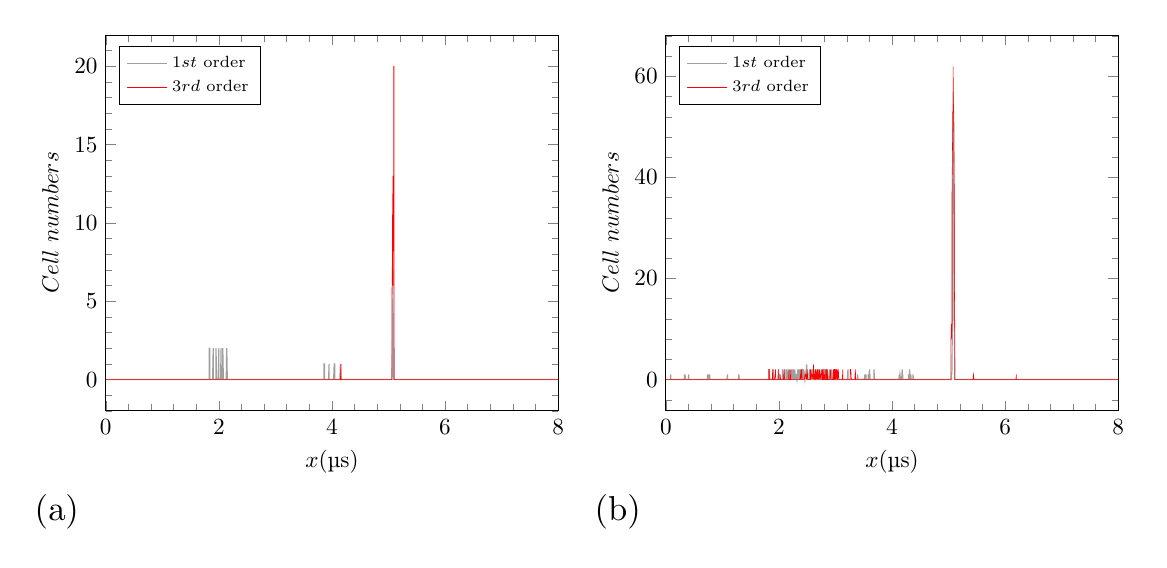}
	\caption{The number of cells where the 1st order and 3rd order scheme used in MOOD-type positivity preserving, $x$ direction reconstrucion: (a) the present WENO-IS scheme; (b) the classical WENO-JS scheme.}
	\label{figs:2d-air-water-bubble-mood-counter-x}
\end{figure}

\begin{figure}[tbh]
\includegraphics[width=\textwidth]{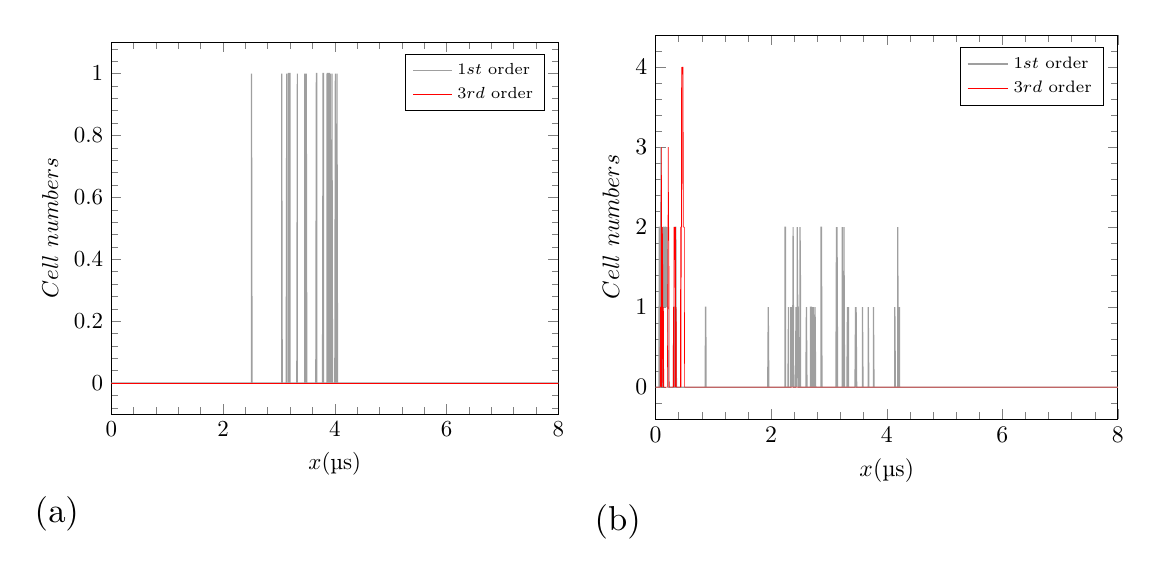}
	\caption{The number of cells where the 1st order and 3rd order scheme used in MOOD-type positivity preserving, $y$ direction reconstrucion: (a) the present WENO-IS scheme; (b) the classical WENO-JS scheme.}
	\label{figs:2d-air-water-bubble-mood-counter-y}
\end{figure}

The numerical schlieren using the classical WENO-JS scheme and the hybrid WENO-MUSCL scheme are also shown in Fig. \ref{figs:2d-air-water-bubble-weno-js-muscl}. The time instance is $t=5.4$ \textmu s, which correspond to the time that the upstream bubble wall impacts on the downstream bubble wall as shown in Fig. \ref{figs:2d-air-water-bubble}b. It shows that the smeared region of the transvers jet using the WENO-JS scheme is a bit larger than the present WENO-IS scheme. Additionally, WENO-JS scheme uses more cells for MOOD-type preserving in $x$ and $y$ direction reconstruction as shown in Fig. \ref{figs:2d-air-water-bubble-mood-counter-x} and Fig. \ref{figs:2d-air-water-bubble-mood-counter-y}, respectively. For the hybrid WENO-MUSCL scheme, it is observed that the two-phase interface for the transverse jet and the wave structures including shock waves and rarefaction waves in the remaining water region using WENO-MUSCL scheme is much smeared than the WENO-IS scheme, which indicates that it has larger dissipation than the WENO-IS scheme. 
The temporal variation of pressure at three locations is shown in Fig. \ref%
{figs:2d-air-water-convergence}. 
\begin{figure}[tbh]
\includegraphics[width=\textwidth]{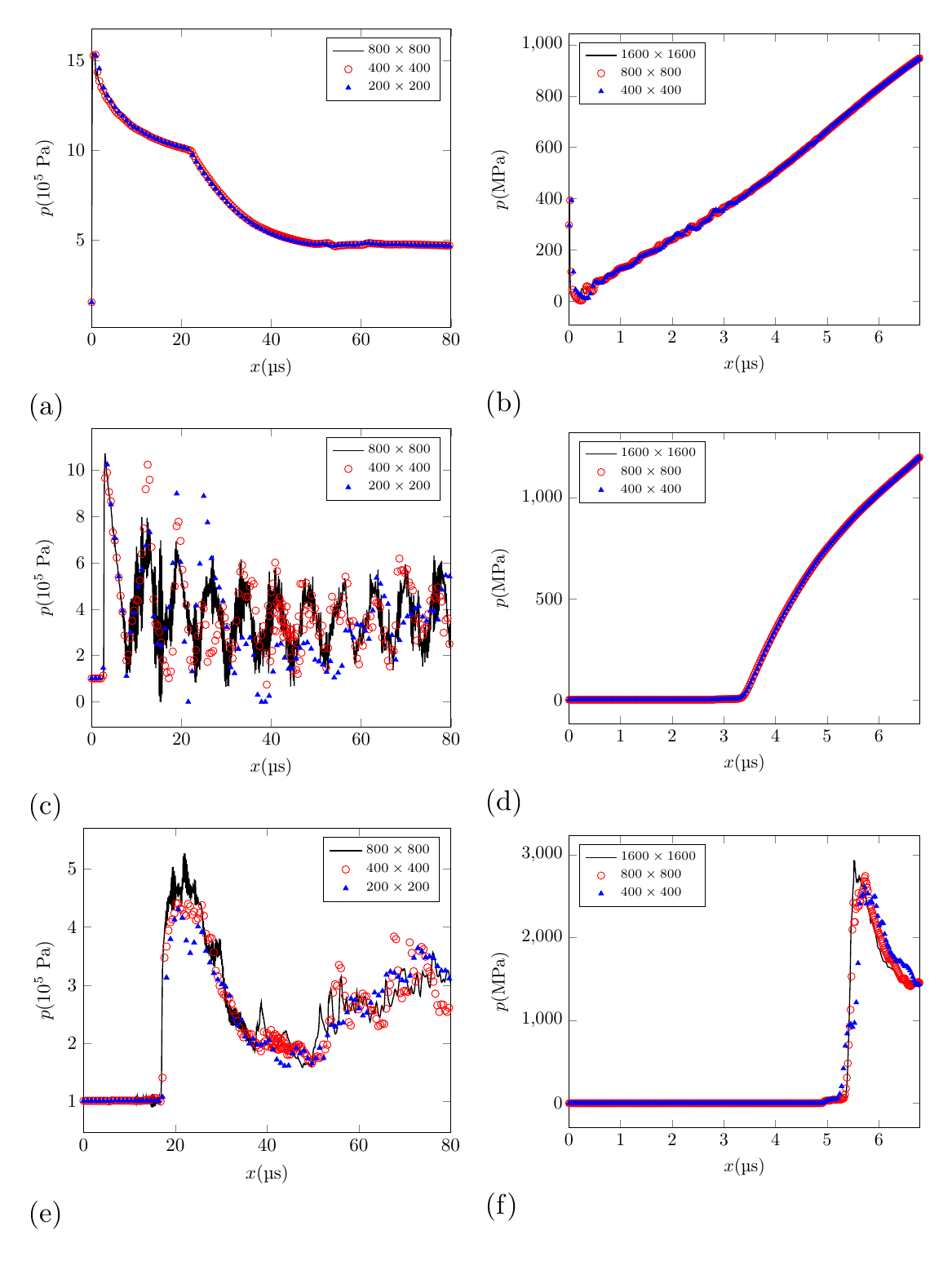}
\caption{The pressure-time profiles: (left) for the shock-droplet case at $%
x_{p}=16$, 12, 8 mm, (right) for the shock-bubble case at $x_{p}=16$, 12, 8
mm.}
\label{figs:2d-air-water-convergence}
\end{figure}
It is observed that the computation results are converged to those on the
finest grid for the shock-bubble case and the early time of the
shock-droplet case. In the later time of the shock-droplet case, no clear
evidence of convergence is found due to the violent interface instability
which leads to the breakup of the droplet. Figure \ref%
{figs:2d-air-water-droplet-convergence-spatial} shows the pressure profiles
along the axial line of the shock-droplet problem at an early-time instance and
the shock-bubble problem at a later-time instance. 
\begin{figure}[tbh]
\includegraphics[width=\textwidth]{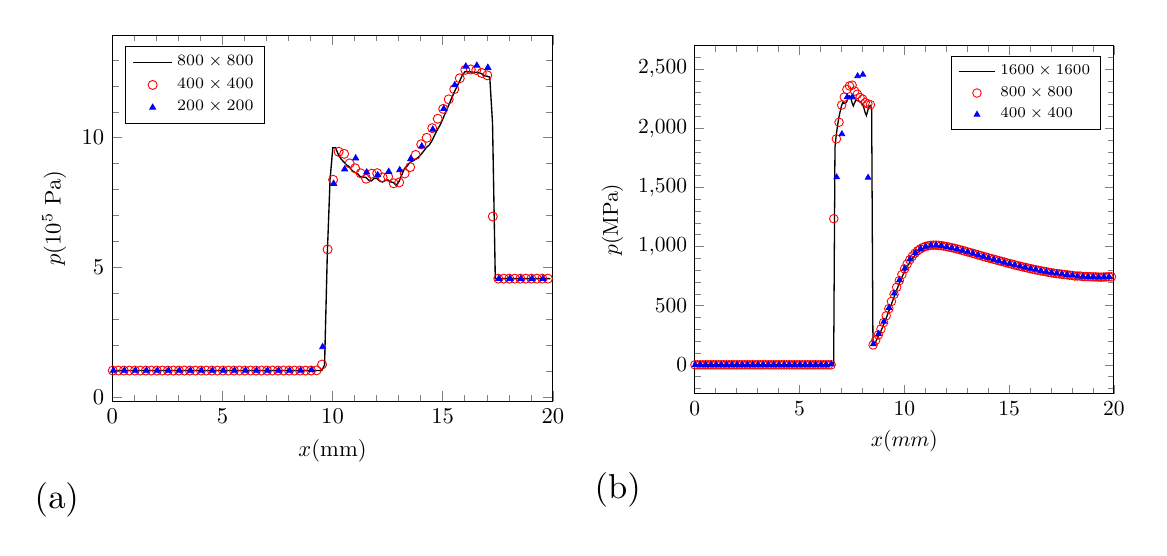}
\caption{The pressure profiles on the horizontal center line at different
time instants: (a) shock droplet case at $t=4.3$ \textmu s, (b)shock
bubble case at $t=5.9$ \textmu s.}
\label{figs:2d-air-water-droplet-convergence-spatial}
\end{figure}
Again the pressure profiles converge to those on the finest grids for both the two-phase problems.

\section{Conclusions}

In this paper, we have developed a 5th-order incremental-stencil WENO
reconstruction method for computing compressible two-phase flow with high
density ratio. Due to the presence of the 2-point candidate stencils, the
present method is able to handle closely located discontinuities, which is a
typical scenario of shock-interface interaction. Furthermore, a MOOD-type
positivity preserving approach is applied to ensure physical meaningful
reconstruction. It is validated with several one-dimensional and
two-dimensional benchmark problems of shock/gas/droplet interactions.
Compared to the hybrid method, the present method is free of
problem-dependent tunable parameters. It also achieves less numerical
dissipation than hybrid method. Note that,
although the present method is based on the quasi-conservative interface
model, it can also be applied to the finite-volume discretization of other
smeared-interface models. A straightforward future work would be
introducing the optimizations used in the target ENO \cite{fu2016family} to
achieve even less numerical dissipation in the smooth region of the
solution. 
\bibliographystyle{unsrt}
\bibliography{incremental-weno}

\begin{thebibliography}{10}

\bibitem{johnsen2009numerical}
E.~Johnsen and T.~Colonius.
\newblock Numerical simulations of non-spherical bubble collapse.
\newblock {\em J. Fluid Mech.}, 629(1):231--262, 2009.

\bibitem{lauer2012numerical}
E.~Lauer, X.Y. Hu, S.~Hickel, and N.A. Adams.
\newblock Numerical modelling and investigation of symmetric and asymmetric
  cavitation bubble dynamics.
\newblock {\em Computers \& Fluids}, 69:1--19, 2012.

\bibitem{dong2009numerical}
G.~Dong, B.~Fan, M.~Gui, and B.~Li.
\newblock Numerical simulations of interactions between a flame bubble with an
  incident shock wave and its focusing wave.
\newblock {\em Proceedings of the Institution of Mechanical Engineers, Part C:
  Journal of Mechanical Engineering Science}, 223(10):2357--2367, 2009.

\bibitem{hirt1981volume}
C.W. Hirt and B.D. Nichols.
\newblock Volume of fluid (vof) method for the dynamics of free boundaries.
\newblock {\em J. Comput. Phys.}, 39(1):201--225, 1981.

\bibitem{rider1998reconstructing}
W.J. Rider and D.B. Kothe.
\newblock Reconstructing volume tracking.
\newblock {\em J. Comput. Phys.}, 141(2):112--152, 1998.

\bibitem{Fedkiw1999}
R.P. Fedkiw, T.~Aslam, B.~Merriman, and S.~Osher.
\newblock A non-oscillatory eulerian approach to interfaces in multimaterial
  flows (the ghost fluid method).
\newblock {\em J. Comput. Phys.}, 152(2):457--492, 1999.

\bibitem{Hu2006}
X.Y. Hu, B.C. Khoo, N.A. Adams, and F.L. Huang.
\newblock A conservative interface method for compressible flows.
\newblock {\em J. Comput. Phys.}, 219(2):553--578, 2006.

\bibitem{niederhaus2008computational}
J.H. Niederhaus, J.A. Greenough, J.G. Oakley, D.~Ranjan, M.H. Anderson, and
  R.~Bonazza.
\newblock A computational parameter study for the three-dimensional
  shock--bubble interaction.
\newblock {\em J. Fluid Mech.}, 594:85--124, 2008.

\bibitem{So2012}
K.K. So, X.Y. Hu, and N.A. Adams.
\newblock Anti-diffusion interface sharpening technique for two-phase
  compressible flow simulations.
\newblock {\em J. Comput. Phys.}, 231(11):4304--4323, 2012.

\bibitem{ansari2013numerical}
M.R. Ansari and A.~Daramizadeh.
\newblock Numerical simulation of compressible two-phase flow using a diffuse
  interface method.
\newblock {\em International Journal of Heat and Fluid Flow}, 42:209--223,
  2013.

\bibitem{coralic2014finite}
V.~Coralic and T.~Colonius.
\newblock Finite-volume weno scheme for viscous compressible multicomponent
  flows.
\newblock {\em J. Comput. Phys.}, 274:95--121, 2014.

\bibitem{abgrall1996prevent}
R.~Abgrall.
\newblock How to prevent pressure oscillations in multicomponent flow
  calculations: a quasi conservative approach.
\newblock {\em J. Comput. Phys.}, 125(1):150--160, 1996.

\bibitem{saurel1999simple}
R.~Saurel and R.~Abgrall.
\newblock A simple method for compressible multifluid flows.
\newblock {\em SIAM Journal on Scientific Computing}, 21(3):1115--1145, 1999.

\bibitem{abgrall2001computations}
R.~Abgrall and S.~Karni.
\newblock Computations of compressible multifluids.
\newblock {\em J. Comput. Phys.}, 169(2):594--623, 2001.

\bibitem{shyue1998efficient}
K.M. Shyue.
\newblock An efficient shock-capturing algorithm for compressible
  multicomponent problems.
\newblock {\em J. Comput. Phys.}, 142(1):208--242, 1998.

\bibitem{shyue1999fluid}
K.M. Shyue.
\newblock A fluid-mixture type algorithm for compressible multicomponent flow
  with van der waals equation of state.
\newblock {\em J. Comput. Phys.}, 156(1):43--88, 1999.

\bibitem{shyue2001fluid}
K.M. Shyue.
\newblock A fluid-mixture type algorithm for compressible multicomponent flow
  with mie--gr{\"u}neisen equation of state.
\newblock {\em J. Comput. Phys.}, 171(2):678--707, 2001.

\bibitem{johnsen2006implementation}
E.~Johnsen and T.~Colonius.
\newblock Implementation of weno schemes in compressible multicomponent flow
  problems.
\newblock {\em J. Comput. Phys.}, 219(2):715--732, 2006.

\bibitem{titarev2004finite}
V.A. Titarev and E.F. Toro.
\newblock Finite-volume weno schemes for three-dimensional conservation laws.
\newblock {\em J. Comput. Phys.}, 201(1):238--260, 2004.

\bibitem{beig2015maintaining}
S.~A. Beig and E.~Johnsen.
\newblock Maintaining interface equilibrium conditions in compressible
  multiphase flows using interface capturing.
\newblock {\em Journal of Computational Physics}, 302:548--566, 2015.

\bibitem{lasheras1998break}
J.C. Lasheras, E.~Villermaux, and E.J. Hopfinger.
\newblock Break-up and atomization of a round water jet by a high-speed annular
  air jet.
\newblock {\em J. Fluid Mech.}, 357:351--379, 1998.

\bibitem{Jiang1996}
G.S. Jiang and C.W. Shu.
\newblock Efficient implementation of weighted eno schemes.
\newblock {\em J. Comput. Phys.}, 126:202--228, 1996.

\bibitem{fu2016family}
L.~Fu, X.Y. Hu, and N.A. Adams.
\newblock A family of high-order targeted eno schemes for compressible-fluid
  simulations.
\newblock {\em J. Comput. Phys.}, 305:333--359, 2016.

\bibitem{clain2011high}
S.~Clain, S.~Diot, and R.~Loubere.
\newblock A high-order finite volume method for systems of conservation
  laws—multi-dimensional optimal order detection (mood).
\newblock {\em J. Comput. Phys.}, 230(10):4028--4050, 2011.

\bibitem{Shu1988}
C.W. Shu and S.~Osher.
\newblock Efficient implementation of essentially non-oscillatory
  shock-capturing schemes.
\newblock {\em J. Comput. Phys.}, 77(2):439--471, 1988.

\bibitem{borges2008improved}
R.~Borges, M.~Carmona, B.~Costa, and W.~S. Don.
\newblock An improved weighted essentially non-oscillatory scheme for
  hyperbolic conservation laws.
\newblock {\em J. Comput. Phys.}, 227(6):3191--3211, 2008.

\bibitem{hu2010adaptive}
X.Y. Hu, Q.~Wang, and N.A. Adams.
\newblock An adaptive central-upwind weighted essentially non-oscillatory
  scheme.
\newblock {\em J. Comput. Phys.}, 229(23):8952--8965, 2010.

\bibitem{einfeldt1991godunov}
B.~Einfeldt, C.D. Munz, P.L. Roe, and B.~Sj{\"o}green.
\newblock On godunov-type methods near low densities.
\newblock {\em J. Comput. Phys.}, 92(2):273--295, 1991.

\bibitem{hu2013positivity}
X.Y. Hu, N.A. Adams, and C.W. Shu.
\newblock Positivity-preserving method for high-order conservative schemes
  solving compressible euler equations.
\newblock {\em J. Comput. Phys.}, 242:169--180, 2013.

\bibitem{zhang2012positivity}
X.X. Zhang and C.W. Shu.
\newblock Positivity-preserving high order finite difference weno schemes for
  compressible euler equations.
\newblock {\em Journal of Computational Physics}, 231(5):2245--2258, 2012.

\bibitem{chen2008flow}
H.~Chen and S.M. Liang.
\newblock Flow visualization of shock/water column interactions.
\newblock {\em Shock Waves}, 17(5):309--321, 2008.

\bibitem{haas1987interaction}
J.F. Haas and B.~Sturtevant.
\newblock Interaction of weak shock waves with cylindrical and spherical gas
  inhomogeneities.
\newblock {\em J. Fluid Mech.}, 181:41--76, 1987.

\bibitem{quirk1996dynamics}
J.~J. Quirk and S.~Karni.
\newblock On the dynamics of a shock--bubble interaction.
\newblock {\em Journal of Fluid Mechanics}, 318:129--163, 1996.

\bibitem{chang2007robust}
C.H. Chang and M.S. Liou.
\newblock A robust and accurate approach to computing compressible multiphase
  flow: Stratified flow model and ausm+-up scheme.
\newblock {\em J. Comput. Phys.}, 225(1):840--873, 2007.

\bibitem{Hu2009}
X.Y. Hu, N.A. Adams, and G.~Iaccarino.
\newblock {On the HLLC Riemann solver for interface interaction in compressible
  multi-fluid flow}.
\newblock {\em J. Comput. Phys.}, 228(17):6572--6589, 2009.

\bibitem{hawker2012interaction}
N.A. Hawker and Y.~Ventikos.
\newblock Interaction of a strong shockwave with a gas bubble in a liquid
  medium: a numerical study.
\newblock {\em J. Fluid Mech.}, 701:59--97, 2012.

\bibitem{wierzba1988experimental}
A.~Wierzba and K.~Takayama.
\newblock Experimental investigation of the aerodynamic breakup of liquid
  drops.
\newblock {\em AIAA Journal}, 26(11):1329--1335, 1988.

\bibitem{igra2001numerical}
D.~Igra and K.~Takayama.
\newblock Numerical simulation of shock wave interaction with a water column.
\newblock {\em Shock Waves}, 11(3):219--228, 2001.

\end{thebibliography}

\end{document}